\documentclass[11pt]{article}
\begin{document}
\def\be{\begin{equation}}
\def\ee{\end{equation}}
                         \def\bearr{\begin{eqnarray}}
                         \def\eearr{\end{eqnarray}}
\def\benum{\begin{enumerate}}
\def\eenum{\end{enumerate}}
\def\bitem{\begin{itemize}}
\def\eitem{\end{itemize}}
                         \def\eg{ {\em e.g.}~}
                         \def\etal{ {\em et al.}~}
                         \def\ie{ {\em i.e.}~}
                         \def\viz{ {\em viz.}~}

\def\lsim{\:\raisebox{-0.5ex}{$\stackrel{\textstyle<}{\sim}$}\:}
\def\gsim{\:\raisebox{-0.5ex}{$\stackrel{\textstyle>}{\sim}$}\:}
\def\go{\rightarrow}
\def\goes{\longrightarrow}
\def\hrar{\hookrightarrow}
\def\bul{\bullet}
\def\mET{E_T \hspace{-1.1em}/\;\:}
\def\mpT{p_T \hspace{-1em}/\;\:}
\def\rpv{$R_p \hspace{-1em}/\;\:$}
\def\rp{$R$-parity}
\def\tb{\tan \beta}
                         \def\N0{\widetilde \chi^0}
                         \def\Cp{\widetilde \chi^+}
                         \def\Cm{\widetilde \chi^-}
                         \def\Cpm{\widetilde \chi^\pm}
                         \def\Cmp{\widetilde \chi^\mp}
\def\l{\lambda}
\def\lp{\lambda'}
\def\lpp{\lambda''}
                         \def\sq{\widetilde q}
                         \def\su{\widetilde u}
                         \def\sd{\widetilde d}
                         \def\sc{\widetilde c}
                         \def\ss{\widetilde s}
                         \def\st{\widetilde t}
                         \def\sb{\widetilde b}
\def\sl{\widetilde \ell}
\def\se{\widetilde e}
\def\snu{\widetilde \nu}
\def\smu{\widetilde \mu}
\def\stau{\widetilde \tau}
\def\tm{\widetilde m}
\def\epem{e^+ e^-}
\thispagestyle{empty}
\begin{flushright}
                                                   TIFR-TH/99-12  \\
                                                   TIFR-HECR-99-01 \\
                                                   hep-ph/9904233
\end{flushright}

\vskip 25pt

\begin{center}

{\LARGE\bf    
        Signals for R-parity-violating Supersymmetry \\ \bigskip
        at a 500 GeV ${\bf e}^{\bf +}{\bf e}^{\bf -}$ Collider } \\

\vskip 25pt

{\bf                         Dilip Kumar Ghosh } \\ 

{\footnotesize\rm 
                     Department of Theoretical Physics, 
                     Tata Institute of Fundamental Research, \\
                     Homi Bhabha Road, Mumbai 400 005, India. \\ 
                     E-mail: dghosh@theory.tifr.res.in  } \\

\bigskip

{\bf                        Rohini M. Godbole } \\ 

{\footnotesize\rm 
                      Centre for Theoretical Studies, 
                       Indian Institute of Science, \\
                        Bangalore 560 012, India. \\ 
                     E-mail: rohini@cts.iisc.ernet.in  } \\ 

\bigskip

{\bf                       Sreerup Raychaudhuri\footnote{
Address after May 1, 1999: \\
\hspace*{0.2in} Department of Physics, Indian Institute of Technology,
Kanpur 208 016, India.}
 } \\ 

{\footnotesize\rm 
                        Department of High Energy Physics,  
                    Tata Institute of Fundamental Research, \\ 
                    Homi Bhabha Road, Mumbai 400 005, India. \\ 
                     E-mail: sreerup@iris.hecr.tifr.res.in  } \\

\vskip 30pt

{\bf                             Abstract 
}

\end{center}

\begin{quotation}
\noindent
We investigate the production of charginos and neutralinos at a 500
GeV $e^+e^-$ collider (NLC) and study their decays to the lightest
neutralino, which then decays into multi-fermion final states through
couplings which do not conserve $R$-parity. These couplings are 
assumed to affect only the decay of the lightest neutralino. 
Detailed analyses of the 
possible signals and backgrounds are performed for five selected points 
in the parameter space.
\end{quotation}

\vskip 60pt

\begin{flushleft}
April 1999 
\end{flushleft}


\newpage
\begin{center}
{\Large\bf 
            1. Introduction: Non-conservation of R-Parity 
}
\end{center}

The Minimal Supersymmetric Standard Model (MSSM) is currently the
most popular option\cite{Rev_SUSY}, for physics beyond the Standard Model 
(SM). A large part of contemporary effort in the search for new physics has,
therefore, been devoted to searches for supersymmetry\cite{SUSY_search}.
One of the cornerstones of most of these search strategies is
the large missing energy and momentum signatures generated by an
undetected neutralino, believed to be the lightest supersymmetric
particle (LSP). This is constrained to be stable and weakly-interacting
with matter because of conservation of a quantum number called \rp, which 
is defined\cite{Farrar-Fayet} to be
\be
                         R = (-1)^{3B+L+2S}
    \label{Rdefn}
\ee
where $B,L$ and $S$ represent, respectively, the baryon number, 
lepton number and intrinsic spin of the particle. This corresponds to
the introduction, by hand, of a discrete $Z_2$ symmetry in the MSSM
Lagrangian. Eq.~(\ref{Rdefn}) implies that $R = 1$ for all SM particles 
and $R = -1$ for all their supersymmetric partners (sparticles). In the SM,
\rp~ is trivially conserved, since each of $B,L$ and $S$ are
separately conserved. The immediate consequence of
\rp~conservation in the MSSM is that sparticles appear in pairs at
each interaction vertex.  Thus, each sparticle decays into another
sparticle while, as mentioned above, the LSP cannot decay at 
all. Moreover the LSP must interact with
matter through exchange of other (off-shell) sparticles.  Since all
the sparticles are known to be heavy, with masses of the
order of the electroweak scale, the LSP's interaction with
matter must be of weak strength; consequently, like the
neutrino, it should escape most detectors. This leads to the
missing energy and momentum signatures mentioned above. Such a 
stable particle is also a prime dark matter candidate\cite{SUSY_darkmatter} 
and, hence, must be
neutral and weakly-interacting. The best candidate for the LSP in the
MSSM is the lightest neutralino, and most experimental
searches\cite{SUSY_search} for supersymmetry assume that this is 
indeed the case.

Since the conservation of \rp~plays such an important role in the
search for supersymmetry, it is important to examine the reasons why
it is believed to be a conserved quantum number. To see this,
we note that the most general superpotential consistent with the
gauge symmetry of the SM has the form\cite{R-parity-violation} 
\bearr
{\cal W} & = &
  \mu            {\hat H}_2  {\hat H}_1
+ h^\ell_{jk}    {\hat H}_2  {\hat L}_j   {\hat{\bar E}}_k
+ h^u_{jk}       {\hat H}_2  {\hat Q}_j   {\hat{\bar U}}_k
+ h^d_{jk}       {\hat H}_1  {\hat Q}_j   {\hat{\bar D}}_k
    \nonumber \\
        &   & \hspace*{-0.16in}
+ \kappa_i       {\hat L}_i  {\hat H}_1 
+ \l_{ijk}       {\hat L}_i  {\hat L}_j   {\hat{\bar E}}_k
+ \lp_{ijk}      {\hat L}_i  {\hat Q}_j   {\hat{\bar U}}_k
+ \lpp_{ijk}     {\hat{\bar U}}_i {\hat{\bar D}}_j  {\hat{\bar D}}_k ,
    \label{superpot}
\eearr
where ${\hat H}_1,{\hat H}_2$ denote $SU(2)$ doublet Higgs
superfields and ${\hat L}$ (${\hat Q}$) denote doublet lepton
(quark) superfields containing the left-chiral leptons (quarks) as
their fermion components.  The ${\hat{\bar E}},{\hat{\bar
U}},{\hat{\bar D}}$ are the $SU(2)$ singlet lepton and quark
superfields containing the right-chiral (charged) anti-leptons and
anti-quarks, while $i,j,k$ are flavour indices. In writing the above, we
have dropped gauge indices, which ensure that ($a$) the $\l_{ijk}$ are
antisymmetric in $i$ and $j$, and ($b$) the $\lpp_{ijk}$ are
antisymmetric in $j$ and $k$. The first three terms in the
second line of Eq.~(\ref{superpot}) can be obtained simply by
replacing, in the previous line, the Higgs superfield doublet ${\hat
H}_2$ by any one of the three lepton superfield doublets ${\hat
L}_i$, a procedure which is made possible by the fact that these
have the same gauge quantum numbers. The last term is a
product of three $SU(2)$ singlets and clearly conserves charge. 

It is immediately apparent that the first three terms in the
second line of Eq.~(\ref{superpot}) violate lepton number ($L$), while
the last violates baryon number ($B$). Both of these are conserved
in the SM. Presence of {\em all} these terms can lead to
catastrophically high rates for proton decay, which are certainly
ruled out. In order to get a proton lifetime consistent with
the current lower \cite{proton_decay} 
bound ($\sim 10^{40}$~s), we would require\cite{Goity-Sher,smirnov_vissani},
typically $\lp \lpp \lsim 10^{-26}$.  Now this is
highly unnatural, unless one or both of $\lp$ and $\lpp$ happen to
be identically zero. The imposition of \rp~conservation ensures
this by forbidding {\em all} the terms in the
second line of Eq.~(\ref{superpot}) and, therefore, constitutes a
simple solution to the problem of fast proton decay in supersymmetric
models.

Though the \rp~argument is rather attractive, it was
pointed out long ago \cite{R-parity-violation} that, as a solution to the proton
decay problem, it constrains the model more than what is really necessary. 
To ensure
vanishing contributions to proton decay, it is quite adequate to
ensure that {\em one} of the couplings in the product $\lp \lpp$
vanishes --- not both. For example\cite{smirnov_vissani}, if we demand baryon number
conservation, and allow lepton number to be violated, then the
$\lpp$ terms vanish and the proton remains stable, but \rp~is no
longer a good symmetry because of the lepton-number-violating terms
in the superpotential. Conversely, if we demand lepton number
conservation, then the first three terms in the second line of
Eq.~(\ref{superpot}) vanish, but the last term, which violates baryon
number, remains as a source of \rp~non-conservation.  We note, in
this context, that the imposition of baryon number conservation is
somewhat better motivated from theoretical considerations \cite{Ibanez-Ross}
than its counterpart in the lepton sector.

It is at once obvious that if we allow for either lepton number or
baryon number to be violated in supersymmetric interactions, then it
should be possible to see some effects in low-energy interactions or
in precision electroweak measurements. Since these have not been
seen to date, it has been shown that all \rp-violating effects at low and
electroweak scale energies must be rather small, and, therefore,
\rp~does survive, at least as an approximate symmetry. However, smallness
of these deviations from the SM is partly ensured by the fact that they 
must occur through the exchange of heavy virtual sfermions and are, thus, 
naturally suppressed.
Despite this, current measurements are still precise enough to constrain
the \rp-violating {\rm couplings} to be rather small, unless,
indeed, the sfermions are very heavy ($\sim 1$ TeV or more). In
fact, there exists a whole series of upper bounds on these couplings
\cite{Dreiner} and some of these are updated regularly in the
literature \cite{French-group,Bhat}.

The situation changes rather dramatically when one considers signals
for supersymmetry at high energy colliders. If we allow for the
non-conservation of \rp, then we have the following possibilities:
\bitem 
\item
The LSP need no longer be stable. It is possible, given a specific structure 
for the \rp-non-conserving sector, to write down its decay modes.
\item
The lightest neutralino need not be the LSP since the arguments in
favour of this are based on the LSP being stable and forming the
major component of cosmic dark matter.
\eitem 
An unstable LSP thus leads to one of the less attractive features of
supersymmetric 
models without \rp~conservation, which is the loss of an excellent
dark matter candidate. However, one can always assume that there are
other candidates such as, for example, invisible axions, machos or wimps.  In
fact, though the existence of certain amounts of galactic dark
matter is undoubted, it might be relevant to point out that the need
for cosmic dark matter is itself based on a theoretical prejudice
that the universe should be closed. The cosmological argument
for conservation of \rp, though plausible, should not be regarded as
a clinching one.

The above discussion makes it clear that there is no compelling
phenomenological reason for the introduction of \rp~as a discrete
symmetry in the MSSM. If, indeed, it is conserved, as the smallness
of upper bounds on \rp-violating couplings may lead us
to suspect, there must be a deep and hitherto unknown theoretical
reason for the corresponding discrete symmetry to be obeyed. In
fact, even the smallness of the couplings is a constraint which is
valid\cite{Dreiner} only if the sfermions are rather light.  
For heavy sfermions,
it is quite possible for \rp~ violation --- in either of its avatars
as lepton number or baryon number violation --- to be quite significant.
If this turns out to be the case, supersymmetric models with R--parity 
violation could become a subject of considerable interest.

It is useful, at this stage, to note that the $\kappa_i$ terms can,
in principle, 
be removed by a redefinition of the lepton doublets ${\hat L}_i$
which leads to absorption of the $\kappa_i$ in the $\l,\lp$
couplings and in the parameters of the scalar potential of the
model. However, these would then reappear at a different energy
scale\cite{Joshipura}. 
Bilinear terms could also lead to a possible vacuum expectation 
value (VEV) for
the sneutrino(s) and mixing of ($a$) charged leptons with
charginos, ($b$) sleptons with charged Higgs bosons, ($c$) neutrinos with
neutralinos and ($d$) sneutrinos with neutral Higgs bosons.
The phenomenological consequences of a sneutrino VEV and
lepton-number-violating mixing have been discussed in the
literature\cite{Hall-Suzuki}, but will not be pursued further in this article.
We are, therefore, making the assumption that the $\kappa_i$ terms are
small. 

Having re-examined the reasons for assuming \rp~conservation and
hence the existence of a stable LSP giving rise to missing energy and 
momentum signals at colliders, we thus come to the conclusion that this is 
merely {\it one} of the possible supersymmetric scenarios and it is certainly 
necessary to make a re-evaluation of supersymmetry signals in the case when the 
LSP {\em can} decay. The implications of \rp~violation for sparticle searches 
have been considered both by theorists and experimentalists. It was 
found~\cite{DPRoy} that the limits on squark and gluino masses from the Tevatron 
data in the presence of \rp~violation are comparable to those obtained
assuming \rp~conservation. Like-sign dileptons can also be used quite 
effectively\cite{DPRoy,Chertok}
for such searches at the Tevatron. Prospects for sparticle searches in the presence of 
\rp~violation at future $pp$ colliders like LHC are also fairly 
bright\cite{French-group}. It was also pointed out~\cite{Roy-Godbole-Tata} that 
chargino and/or neutralino production at LEP would be a good place to look for 
signals for supersymmetry with broken $R$-parity.

Studies of the above nature, neglected until a short time ago, have
now been included in the agenda of most of the major running
experimental facilities, such as those at LEP-2 and the Tevatron.
The four experimental collaborations at LEP have made
separate studies \cite{LEP-Collaborations} of signals for \rp-violating
supersymmetry and we take note of their bounds in our
analysis. Similar bounds exist from the Tevatron Run I data\cite{Tevatron}. 
However, we must note that LEP is already near the end of
its kinematic reach, and unless, indeed, supersymmetric particles are
lying just around the corner, waiting to be discovered, it is
unlikely that we will obtain anything more than improved bounds
from the present and future runs of LEP. Better hopes may be placed
on the Run II of the Tevatron and the TeV-33 run\footnote{Should it happen}, 
and, of course, the LHC, 
but it is really to the clean environment of a 500 GeV linear $\epem$
collider, such as the planned Next Linear Collider (NLC), that we
must look\cite{Peskin-Murayama,Dreiner-Lola} if we wish to complement the information
from hadron colliders and further help find unambiguous signals for low-energy
supersymmetry\footnote{Assuming that the sparticle spectrum is not 
too heavy for such a machine.}.
The present work is devoted therefore, to a preliminary 
study of the production of charginos and neutralinos at a
500 GeV linear $\epem$ collider, and their decays in the
presence of \rp-violating couplings. Signal studies for the
\rp-conserving option may be found in the literature
\cite{MSSM_NLC,NLC-Report}. What really distinguishes our analysis from similar
studies at LEP-2 is the higher collision energy, which opens up the
possibility of copious production of the {\em heavier} {\rm chargino 
and neutralino states}. Their cascade decays can then add to the signals, 
but also introduce additional complications because of the multiplicity of
possible final states. We present here a first attempt to sort out
these possibilities and obtain distinguishable signals. It turns out that the 
heavier chargino/neutralino states,
though they make the analysis complicated, are ultimately more
of a help than a hindrance in signatures deciphering for 
\rp-violating supersymmetry from the backgrounds. 

The particular scenario which is discussed in this article is the
so--called {\em weak limit} of \rp~violation \cite{Baer-Kao-Tata}, where
the \rp-violating couplings are considerably smaller than the gauge
couplings. This is partly motivated by the fact that 
current upper bounds on the \rp-violating couplings are
typically an order of magnitude below the gauge couplings for
sfermion masses near the electroweak symmetry-breaking scale\cite{French-group}.  
However, this is not a particularly compelling argument, 
since we have just argued the case for heavy sfermions relaxing
these bounds. The current choice of scenario should, therefore, be 
regarded as motivated principally by a requirement of 
{\em minimal deviation} from the \rp-conserving model. As a result, most
production and decay processes in this analysis will take place along
the lines expected in the MSSM in the \rp-conserving case.  Though
it is no longer an absolute necessity (irrespective of the strength
of the couplings), the LSP will still be assumed to be the lightest 
neutralino, and its decays will form the chief point of departure of
our study from earlier ones involving a stable and invisible LSP.  
Our study is thus, in a sense, complementary to the work of
Refs.~\cite{Choudhury,Dreiner-Lola,Chemtob-Moreau}, who discuss processes 
at $\epem$ colliders
where the \rp~violation takes place directly at the production
vertices. These studies require large \rp-violating couplings, while
the one undertaken here assumes small couplings. 

Though we assume the \rp-violating couplings to be much smaller
than the gauge couplings, we still require 
the couplings relevant for decay of the LSP to be large enough to
ensure that it decays within the detector. This point is, however, 
somewhat academic, since, even with couplings one or two orders of
magnitude below the current upper bound, the LSP will decay (for all
practical purposes) at the interaction point itself and thus will not
even exhibit displaced vertices.

The plan of this article is as follows. A large part of the material
that goes into a discussion of chargino and neutralino production and 
decay in the MSSM is already available in the literature.  In the following 
section, we make some general remarks concerning processes which lead to 
the production of a pair of charginos or neutralinos at an $\epem$ collider
and present contour plots in the parameter space showing the
importance of each channel at NLC energies. We also explain our
choice of the five points in the parameter space where a detailed
analysis has been done. The results of this section are relevant even if 
\rp~is assumed to be conserved. In section 3 we discuss the decay of the 
LSP into multi-fermion channels through \rp-violating couplings and then
go on to consider decay modes of the heavier charginos and neutralinos to the
LSP (which are again of general interest). Sections 4, 5 and 6 are devoted 
to a study of the different signals predicted in the case of $\l,\lp,\lpp$
couplings respectively as well as the relevant backgrounds. 
Our results are summed-up in Section 7.
In the Appendices, we present detailed formulae for the decay of the
LSP and exhibit typical results which might be of interest if the
NLC has a $\sqrt{s} = 350$ GeV centre-of-mass stage. We also 
discuss some details of the background studies. 


\vspace{0.3in}
\begin{center}
{\Large\bf 
         2. Chargino and Neutralino Pair-production in 
             ${\bf e}^{\bf +}{\bf e}^{\bf -}$ Collisions
} \end{center}

Computing production cross sections for charginos and neutralinos 
in the MSSM is a complex business. There are no \rp-violating contributions 
to the production mechanism in the weak limit. Nevertheless, it is difficult 
to make remarks which hold in
the general case. The cross sections depend critically on the masses
and mixing angles of charginos and neutralinos, which, in turn, depend on the
parameters $M_1, M_2, \mu, \tb$ of which the mass matrices are made
up. In this article, we assume gaugino mass unification
at a high scale, which means that $M_1$ and $M_2$ are linearly
related by:
\be
M_1 = \frac{5}{3} \tan^2 \theta_W M_2 \ . 
\ee
Thus, the free parameters in question are $M_2, \mu, \tb$
and, of course, the masses of the left and right selectrons. The
mass of the electron sneutrino is not a free parameter, being
related to the mass of the left selectron by the $SU(2)$-breaking
relation
\be
m^2_{\se_L} = m^2_{\snu_e} - \frac{1}{2} M^2_W \cos 2 \beta \ .
     \label{SU2_breaking}
\ee
The free parameters $M_2, \mu, \tb$ decide the masses as well as the mixing
angles of the charginos and neutralinos.
Apart from the mixing angles, which go into the Feynman rules, the other
major question in determining the production cross sections is that 
of kinematics. For a centre-of-mass energy of 500 GeV, it
is rather difficult to produce pairs of the heavier charginos $\Cpm_2$ or
the heavier neutralinos $\N0_3$ and $\N0_4$, except in narrow
regions of the parameter space where both the charginos or all the
neutralinos are light. However, the lighter states, $\Cpm_1, \N0_1, \N0_2$, 
can be freely produced. 

Given the large number of unknown parameters and the variety of 
scenarios for $R$-parity violation, it is not practical to present 
a detailed analysis of the signals over the {\it entire} parameter
space. We have chosen, for detailed analysis,  
five particular points in the MSSM parameter space which are allowed by 
the current LEP data and where the masses and mixings of the gaugino
and higgsino states 
have different, but typical, features\footnote{The inspiration for
this comes from the procedure followed for the constrained MSSM at 
Snowmass\cite{Snowmass}, though we have not taken their precise values}. 
These five points are
\vspace*{-0.45in} 
\begin{center}
$$
\begin{array}{ccccc}
({\bf A}) & M_2 = 100~{\rm GeV}, & \mu = -200~{\rm GeV}, & \tb = ~2, 
    & M_{\se_L} = M_{\se_R} = 150~{\rm GeV}, \\
({\bf B}) & M_2 = 150~{\rm GeV}, & \mu = +150~{\rm GeV}, & \tb = 20,
    & M_{\se_L} = M_{\se_R} = 150~{\rm GeV}, \\
({\bf C}) & M_2 = 150~{\rm GeV}, & \mu = +250~{\rm GeV}, & \tb = ~2,
    & M_{\se_L} = M_{\se_R} = 150~{\rm GeV}, \\
({\bf D}) & M_2 = 150~{\rm GeV}, & \mu = -250~{\rm GeV}, & \tb = 20,
    & M_{\se_L} = M_{\se_R} = 200~{\rm GeV}, \\
({\bf E}) & M_2 = 200~{\rm GeV}, & \mu = -250~{\rm GeV}, & \tb = 20,
    & M_{\se_L} = M_{\se_R} = 200~{\rm GeV}.
\end{array}
$$
\end{center}
\noindent Apart from the obvious fact that round numbers have been
used for $M_2$ and $\mu$, we have concentrated on two values of 
$\tb$, namely 2 and 20. The lower value corresponds to what is
likely to be the lower bound\cite{LEP-Higgs} 
on $\tb$ from non-observation of the lightest Higgs boson
when LEP-2 has finished its run, while the upper value is a reasonably
high one, though not quite the highest allowed by the present constraints.
Throughout this analysis, the soft supersymmetry-breaking squark masses have 
been set to the common value 500 GeV, and the trilinear couplings to
$A_t = A_b = A_\tau = 0$. 

\footnotesize
$$
\begin{array}{|c|cc|cc|}
\hline
{\rm Point} & {\rm Particle} & {\rm mass}
& {\rm wino}~{\tilde W}^\pm & {\rm higgsino}~{\tilde H}^\pm \\
\hline\hline
{\bf A}  & (\Cpm_1)_L   &  112.1~{\rm GeV} & ~~0.0~\%  & 100.0~\%  \\
         & (\Cpm_1)_R   &                  & ~20.5~\%  & ~79.5~\%  \\
         & (\Cpm_2)_L   &  224.3~{\rm GeV} & 100.0~\%  & ~~0.0~\%  \\
         & (\Cpm_2)_R   &                  & ~79.5~\%  & ~20.5~\%  \\
\hline
{\bf B}  & (\Cpm_1)_L   &  ~99.9~{\rm GeV} & ~33.1~\%  & ~66.9~\%  \\
         & (\Cpm_1)_R   &                  & ~66.9~\%  & ~33.1~\%  \\
         & (\Cpm_2)_L   &  218.8~{\rm GeV} & ~66.9~\%  & ~33.1~\%  \\
         & (\Cpm_2)_R   &                  & ~33.1~\%  & ~66.9~\%  \\
\hline
{\bf C}  & (\Cpm_1)_L   &  110.5~{\rm GeV} & ~17.5~\%  & ~82.5~\%  \\
         & (\Cpm_1)_R   &                  & ~28.0~\%  & ~72.0~\%  \\
         & (\Cpm_2)_L   &  292.7~{\rm GeV} & ~82.5~\%  & ~17.5~\%  \\
         & (\Cpm_2)_R   &                  & ~72.0~\%  & ~28.0~\%  \\
\hline
{\bf D}  & (\Cpm_1)_L   &  135.2~{\rm GeV} & ~~6.9~\%  & ~93.1~\%  \\
         & (\Cpm_1)_R   &                  & ~27.8~\%  & ~72.2~\%  \\
         & (\Cpm_2)_L   &  282.1~{\rm GeV} & ~93.1~\%  & ~~6.9~\%  \\
         & (\Cpm_2)_R   &                  & ~72.2~\%  & ~27.8~\%  \\
\hline
{\bf E}  & (\Cpm_1)_L   &  173.4~{\rm GeV} & ~18.0~\%  & ~82.0~\%  \\
         & (\Cpm_1)_R   &                  & ~41.2~\%  & ~58.8~\%  \\
         & (\Cpm_2)_L   &  292.1~{\rm GeV} & ~82.0~\%  & ~18.0~\%  \\
         & (\Cpm_2)_R   &                  & ~58.8~\%  & ~41.2~\%  \\
\hline
\end{array}
$$
\vskip 2pt 
\normalsize
\centerline{ {\bf Table 1.}~~{\it Masses and compositions of the two
charginos at the five selected points (see text).}}
\vskip 5pt 

The masses and compositions of the charginos at these
points are given in Table 1. It is at once obvious that at point
(A), the left-handed charginos are pure states, while the
right-handed charginos are mixed states. At the other points both left- and
right-handed charginos are mixed states, though
point (D) is close to a pure state. The mass of the lighter
chargino varies from about 100 GeV for point (B) to near the
top quark mass for point (E). We thus get a span of most of the different 
possibilities. Note that $\Cpm_2$ has mass above 200 GeV (and 
$ M_{\se_L} > M_{\Cpm_1} $) for all the five points. 

Charginos will be pair-produced in $\epem$ collisions through the
three Feynman diagrams of Fig.~1($a$). Except for the
photon-exchange diagram, the others can result in production of a
pair of dissimilar charginos (provided it is kinematically
allowed). These diagrams have been evaluated before and formulae
are readily available in the literature\cite{Gaugino-production}. 
In this section, we confine
ourselves to some general remarks on the production process.

Noting (see, Table 1) that the chargino is a linear 
combination of a wino and a charged higgsino state, and that its 
couplings are obtained by
supersymmetrizing the corresponding gauge and Yukawa couplings
respectively, it
is obvious that the $t$-channel sneutrino-exchange diagram makes a
significant contribution only if the charginos 
$\Cpm_1$ and $\Cpm_2$ have substantial wino components, since
the coupling of the higgsino components to electrons are suppressed by
$m_e/M_W$~($\sim 10^{-6}$). 
In such regions of the parameter space, however, 
the same $t$-channel diagram interferes
destructively with the others, leading to a well-known 
dip in the cross section
when it is plotted as a function of the mass of the exchanged
electron sneutrino.

\begin{figure}[h]
\begin{center}
\vspace*{3.6in}
      \relax\noindent\hskip -6.0in\relax{\includegraphics{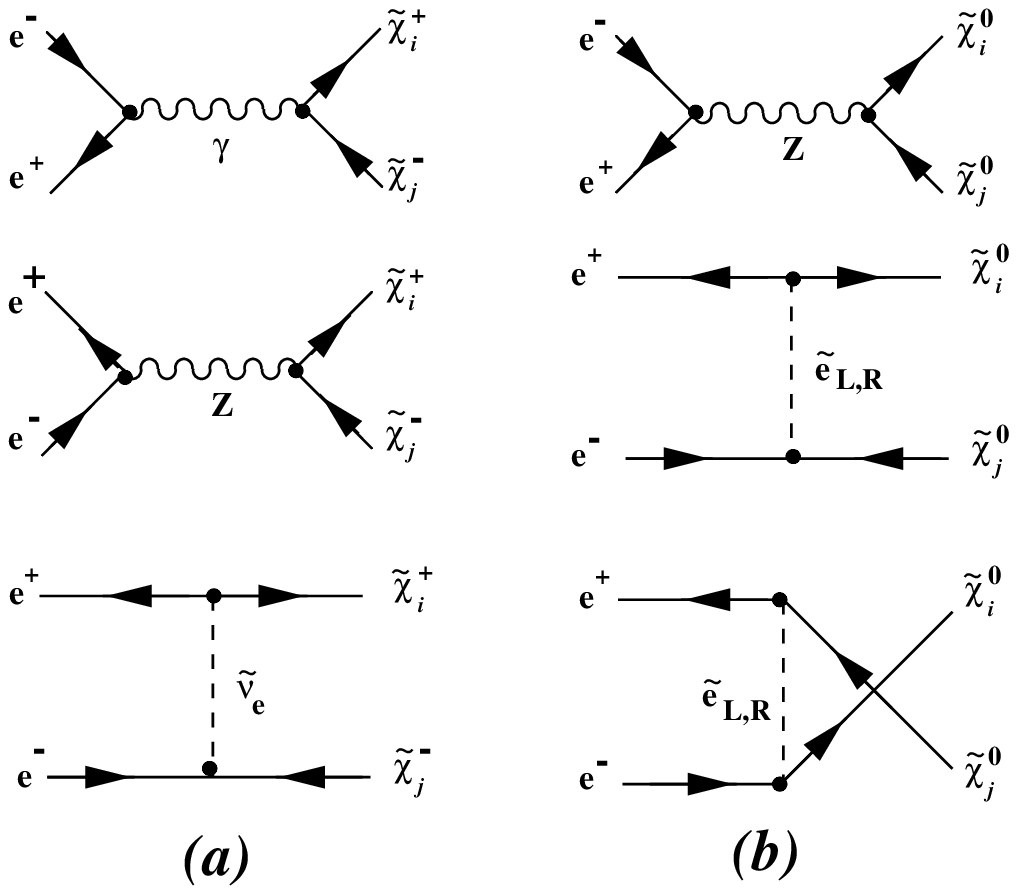}}
\end{center}
\end{figure}
\vspace*{-0.30in}
\noindent
{\bf Figure 1.}~{\it  
Feynman diagrams contributing to the processes 
($a$)~$\epem~\go~\Cp_i~\Cm_j$ and 
($b$)~$\epem~\go~\N0_i~\N0_j$ in the MSSM.}
\vspace{0.1in}

In Fig.~2($A$) and 2($B$), we plot the cross sections for production
of 
$$\Cp_1 \Cm_1, \qquad \Cpm_1 \Cmp_2, \qquad \Cp_2 \Cm_2,$$ 
as functions of
the mass of the left selectron, which is related 
to the mass of the electron-sneutrino by Eq.~(\ref{SU2_breaking}). 
Fig.~2($A$) corresponds to the point (A) and Fig.~2($B$) corresponds to 
the point (B) in the list of selected points
in the parameter space. One must remember that the cross section
for production of a pair of dissimilar charginos should be
multiplied by a combinatoric factor of 2. It is now obvious
that the cross section for the production of a pair of lighter
charginos $\Cpm_1 \Cmp_1$ is much larger than similar cross
sections for production of the heavier chargino states.
Cross-sections for production of one light and one heavy state get
heavily suppressed as the sneutrino mass grows larger and thus may
be seen to originate almost entirely from the $t$-channel sneutrino
exchange. As a result, the suppression is stronger in case (A) where
the charginos are pure states. Fig.~2($B$), however, shows the characteristic 
dip in production of $\Cp_2 \Cm_2$ due to $s$ and $t$-channel 
interference. It is clear therefore, that the bulk of the signal will originate
from production of a pair of lighter charginos.

\begin{figure}[h]
\begin{center}
\vspace*{1.8in}
      \relax\noindent\hskip -4.0in\relax{\includegraphics{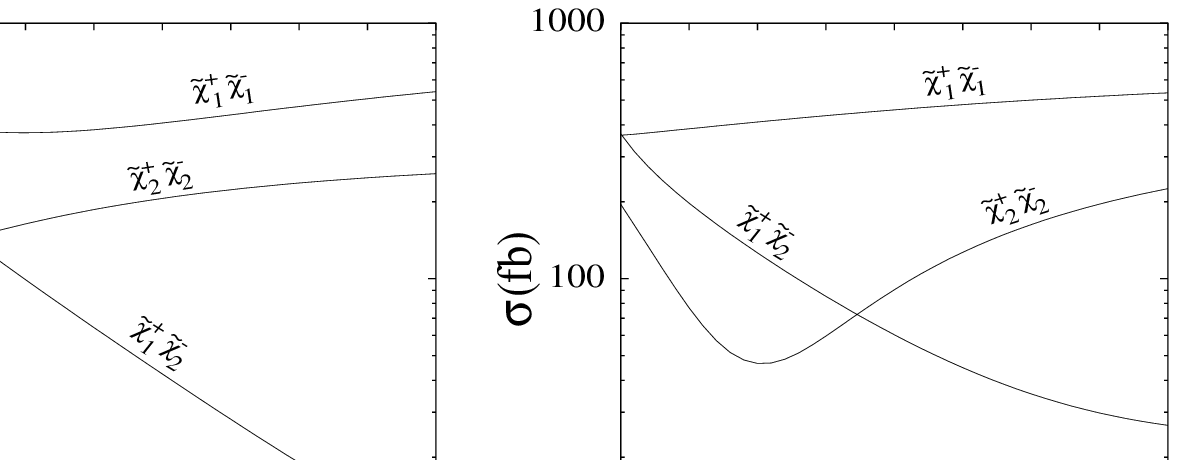}}
\end{center}
\end{figure}
\vspace{0.3in}
\noindent
{\bf Figure 2.}~{\it  Cross-sections for chargino pair
production with parameter choices {\em (A)} and {\em (B)} as given 
in the text. Cross sections for dissimilar charginos should be 
multiplied by a factor of 2. }
\vspace{0.1in}

Neutralinos will also be pair-produced in $\epem$ collisions through the
Feynman diagrams given in Fig.~1($b$). There is (naturally) no
photon-exchange diagram, but now there is a $u$-channel as well as a
$t$-channel diagram with selectron exchange to which both left and
right selectrons can contribute. This is because of the Majorana
nature of the neutralinos. The final amplitude is thus the coherent 
sum of five diagrams with the signs chosen carefully so that $t$-
and $u$-channels add to the $s$-channel diagram with opposite signs 
because of odd and even numbers of permutations arising in
the Wick contractions. All the diagrams can result in the
production of similar or dissimilar neutralinos, so that we can have
ten distinct final states. As in the case of charginos, these
diagrams have been evaluated before and formulae are readily
available in the literature\cite{Gaugino-production}. 

The process $\epem \goes \N0_i \N0_j$, like the analogous process
with charginos, depends crucially on the composition of the
neutralino(s) involved. As in the case of charginos, we may safely
conclude that the higgsino components of the neutralino are
irrelevant. For right selectron exchange, in fact, the wino component
is also irrelevant and the electron-left selectron coupling depends
solely on the bino component.
Thus, the contributions of the $t$ and $u$-channel diagrams in
Fig.~1($b$) require substantial gaugino components of the relevant
neutralinos. For the $s$-channel diagram, it is a different story.
The $Z \N0_i \N0_j$ coupling arises from supersymmetrization of
the $ZH^0_1H^0_2$ coupling and thus a significant contribution can
arise only if there is a significant higgsino component in each of
the two neutralinos. Fortunately, the structure of the
neutralino mass matrix in the higgsino sector is such\cite{Rev_SUSY} 
that a higgsino-dominated neutralino must be a near-equal mixture of the
two higgsino states, so that this scenario covers a greater part of
the parameter space than what might naively be expected.

\footnotesize
$$
\begin{array}{|c|cc|ccc|}
\hline
{\rm Point} & {\rm Particle} & {\rm mass} & {\rm photino}~{\tilde \gamma}
& {\rm zino}~{\tilde Z} & {\rm higgsinos}~{\tilde H}_1^0,~{\tilde H}_2^0\\
\hline\hline
{\bf A} & \N0_1 &  ~54.3~{\rm GeV} &  86.9~\%  &  11.3~\%  &  ~1.8~\%  \\
        & \N0_2 &  111.9~{\rm GeV} &  12.9~\%  &  73.5~\%  &  13.6~\%  \\
        & \N0_3 &  208.5~{\rm GeV} &  ~0.2~\%  &  ~7.9~\%  &  91.9~\%  \\
        & \N0_4 &  224.4~{\rm GeV} &  ~0.0~\%  &  ~7.3~\%  &  92.7~\%  \\
\hline
{\bf B} & \N0_1 &  ~63.4~{\rm GeV} &  40.4~\%  &  35.1~\%  &  24.5~\%  \\
        & \N0_2 &  107.3~{\rm GeV} &  56.5~\%  &  12.1~\%  &  31.4~\%  \\
        & \N0_3 &  163.6~{\rm GeV} &  ~0.1~\%  &  ~4.6~\%  &  95.3~\%  \\
        & \N0_4 &  218.3~{\rm GeV} &  ~3.0~\%  &  48.1~\%  &  48.9~\%  \\
\hline
{\bf C} & \N0_1 &  ~62.5~{\rm GeV} &  46.9~\%  &  43.4~\%  &  ~9.7~\%  \\
        & \N0_2 &  117.7~{\rm GeV} &  52.6~\%  &  33.0~\%  &  14.4~\%  \\
        & \N0_3 &  252.3~{\rm GeV} &  ~0.0~\%  &  ~0.6~\%  &  99.4~\%  \\
        & \N0_4 &  297.6~{\rm GeV} &  ~0.5~\%  &  22.9~\%  &  76.5~\%  \\
\hline
{\bf D} & \N0_1 &  ~73.7~{\rm GeV} &  70.0~\%  &  25.7~\%  &  ~4.3~\%  \\
        & \N0_2 &  135.2~{\rm GeV} &  29.5~\%  &  53.6~\%  &  17.0~\%  \\
        & \N0_3 &  262.0~{\rm GeV} &  ~0.0~\%  &  ~3.0~\%  &  97.0~\%  \\
        & \N0_4 &  278.6~{\rm GeV} &  ~0.5~\%  &  17.7~\%  &  81.8~\%  \\
\hline
{\bf E} & \N0_1 &  ~97.7~{\rm GeV} &  68.5~\%  &  26.3~\%  &  ~5.3~\%  \\
        & \N0_2 &  173.6~{\rm GeV} &  29.6~\%  &  42.1~\%  &  28.3~\%  \\
        & \N0_3 &  260.8~{\rm GeV} &  ~0.0~\%  &  ~2.5~\%  &  97.5~\%  \\
        & \N0_4 &  290.1~{\rm GeV} &  ~1.9~\%  &  29.2~\%  &  68.9~\%  \\
\hline
\end{array}
$$
\vskip 5pt
\normalsize \noindent 
{\bf Table 2.}~{\it Masses and compositions of neutralinos at
the five selected points (see text). In the last column, we
give the combination of the two neutral higgsinos.}
\vspace{0.1in}

Following the same procedure as for charginos, we detail in Table 2 
the composition of the
neutralinos at the five selected points. At all the points,
the LSP is gaugino-dominated, being mainly photino in cases
(A), (D) and (E) and having substantial zino components in cases 
(B)--(E). It has a substantial higgsino component only in case
(B). The higher neutralinos $\N0_3$ and $\N0_4$ have none or very small
photino components, in all the cases. The $\N0_3$ turns out to be
mostly higgsino-dominated, while the $\N0_4$, though principally
higgsino, has substantial zino components as well.

In Fig.~3($a-f$), we plot the cross sections for production of
$\N0_i \N0_j$ as functions of \\
($a,b$) the mass of the left selectron, when the mass of the right 
selectron is set to 150 GeV; \\
($c,d$) the mass of the right selectron, when the mass of the left 
selectron is set to 150 GeV, and \\
($e,f$) the common mass of the selectron in the case when left and 
right selectrons are degenerate. \\
Only the combinations 
$$\N0_1 \N0_1, \qquad 
  \N0_1 \N0_2, \qquad 
  \N0_1 \N0_3, \qquad 
  \N0_1 \N0_4, \qquad 
  \N0_2 \N0_2, \qquad
  \N0_2 \N0_3, \qquad 
  \N0_2 \N0_4
$$ 
have been shown. The graphs on the left, namely ($a,c,e$),
correspond to the point (A), while the graphs on the right, namely
($b,d,f$), correspond to the point (C).  One must also remember that
cross sections with dissimilar neutralinos carry an extra combinatoric
factor of 2, while cross sections with identical neutralinos already
carry a factor of half because of quantum statistics. It is now
obvious from the figure, that the production of $\N0_1$ and $\N0_2$,
either as similar or dissimilar pairs, exceeds production of the
heavier states $\N0_3, \N0_4$ roughly by an order of 
magnitude\footnote{Unless
the selectron is very heavy; in this case, all the cross sections are 
small anyway.}. This
is an important result, and it will enable us, in the next section,
to simplify search strategies considerably.

\begin{figure}[h]
\begin{center}
\vspace*{5.0in}
      \relax\noindent\hskip -5.0in\relax{\includegraphics{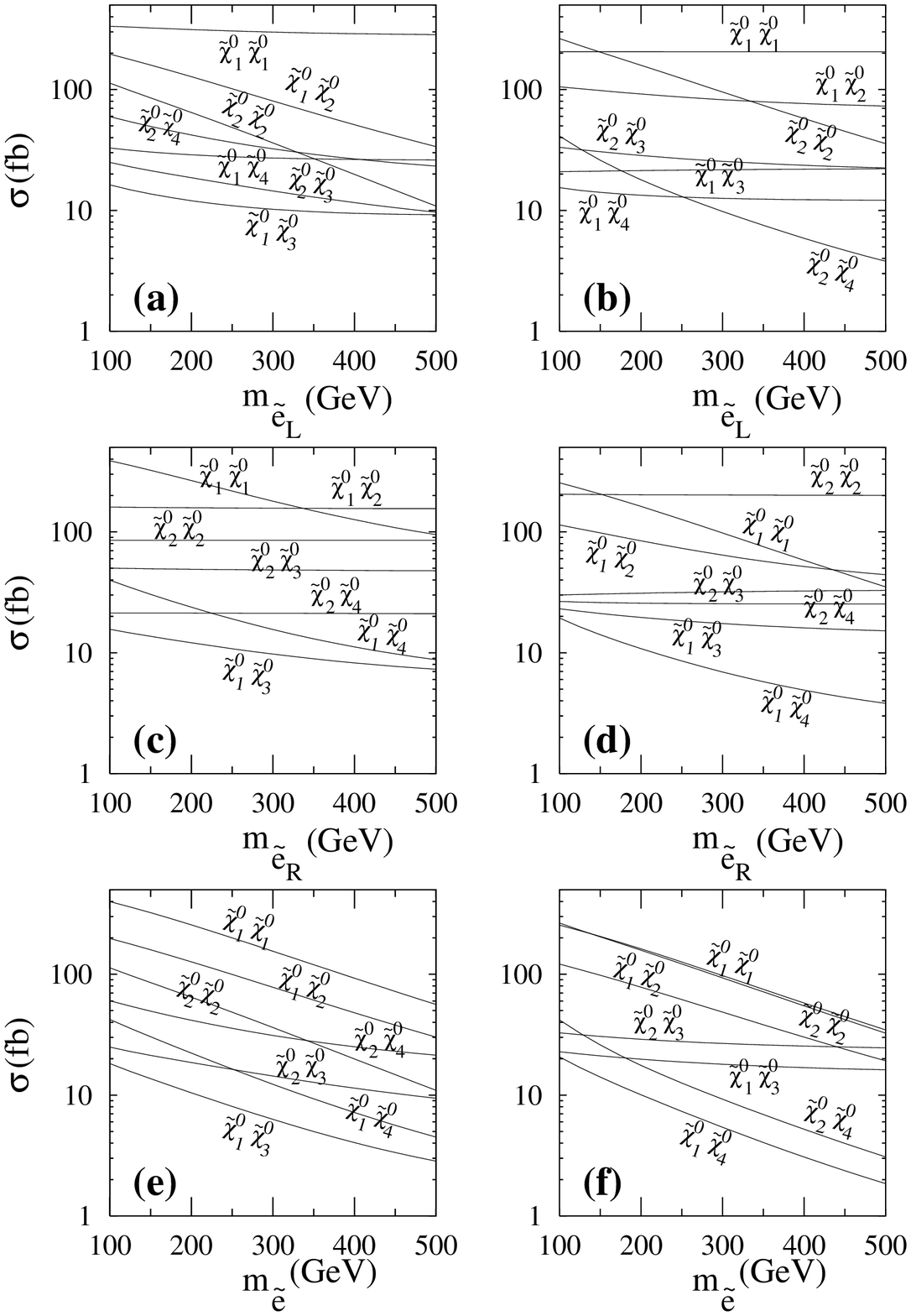}}
\end{center}
\end{figure}
\vspace*{0.8in}
\noindent
{\bf Figure 3.}~{\it Cross-sections for neutralino pair
production at the points {\em (A)} and {\em (C)}. Cross sections for
dissimilar neutralinos should be multiplied by 2. For
($e,f$) left and right selectrons are degenerate. }
\vspace{0.1in}

The illustrative results in Figs.~2 and 3 are, of course, obtained by 
considering three of the selected points in the parameter space. It is
now relevant to ask whether
we can generalize the principal conclusion --- that it is only
important to consider production modes of the $\N0_1$, $\N0_2$ and
$\Cpm_1$ to observe signals at a 500 GeV machine --- to the entire
parameter space. Part of the answer is given by kinematics: the
current bounds\cite{LEP-Collaborations} on the $\Cpm_1 $ mass from LEP-2 
force the heavier
states $\N0_3$, $\N0_4$ and $\Cpm_2$ to have masses fairly
close to the kinematic reach of a 500 GeV machine.  This is
illustrated in Fig.~4, where we have shown scatter plots of the
heavier chargino and neutralino masses as $M_2$ varies over the range 
0 to 600 GeV, $\mu$ varies from -600 to 600 GeV and $\tb$ varies from 1.5 to 50.
Clearly, when there is no experimental constraint on the MSSM,
all the states are well within the kinematic reach of a 500 GeV
machine (which corresponds to the 250 GeV box). Imposition of the LEP-2
constraint $\Cpm_1 > 91$ GeV immediately drives the heaviest
chargino and neutralino states (practically) out of the box, while
the intermediate states $\N0_2$ and $\N0_3$ are driven to heavier,
but not kinematically inaccessible values.  One can therefore,
safely write off the $\N0_4$ and $\Cpm_2$ states. It only remains,
therefore, to ask whether it is relevant to include the $\N0_3$ in
our analysis. For this, the {\it dynamics} of the production cross
section(s) must now be considered.

\begin{figure}[h]
\begin{center}
\vspace*{2.5in}
      \relax\noindent\hskip -5.0in\relax{\includegraphics{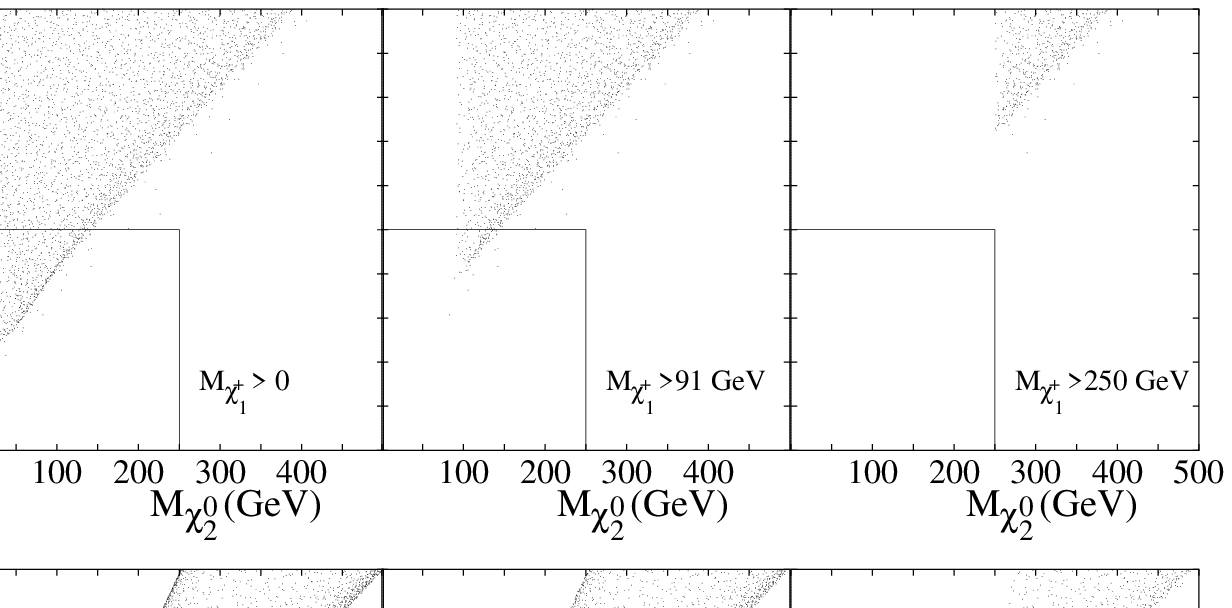}}
\end{center}
\end{figure}
\vspace{1.7in}
\noindent
{\bf Figure 4.}~{\it Scatter plots for higher chargino and neutralino 
masses in the MSSM. The ranges of variation of the parameters are given in
the text. The box represents the kinematic reach of a 500 GeV machine.}

Before we discuss this, however, it is interesting to look at the
graphs on the extreme right of Fig.~4. These would correspond to the
(disappointing) situation when a 500 GeV $\epem$ collider has
completed its run without finding any evidence of supersymmetry and
has been able to push the lower bound on the mass of the lighter
chargino to $\sim $ 250 GeV. It is likely that the machine would then be 
upgraded to 1 TeV in energy\cite{NLC-Report}, 
kinematic limits for which correspond
to the outer boxes in the figure. In this case, we see that the same
pattern repeats itself, with the heavier states being pushed
almost out of the picture. Thus the broad conclusions of this work
would still hold in the newer context and almost identical studies 
can be made. 

\begin{figure}[ht]
\begin{center}
\vspace*{4.2in}
      \relax\noindent\hskip -5.0in\relax{\includegraphics{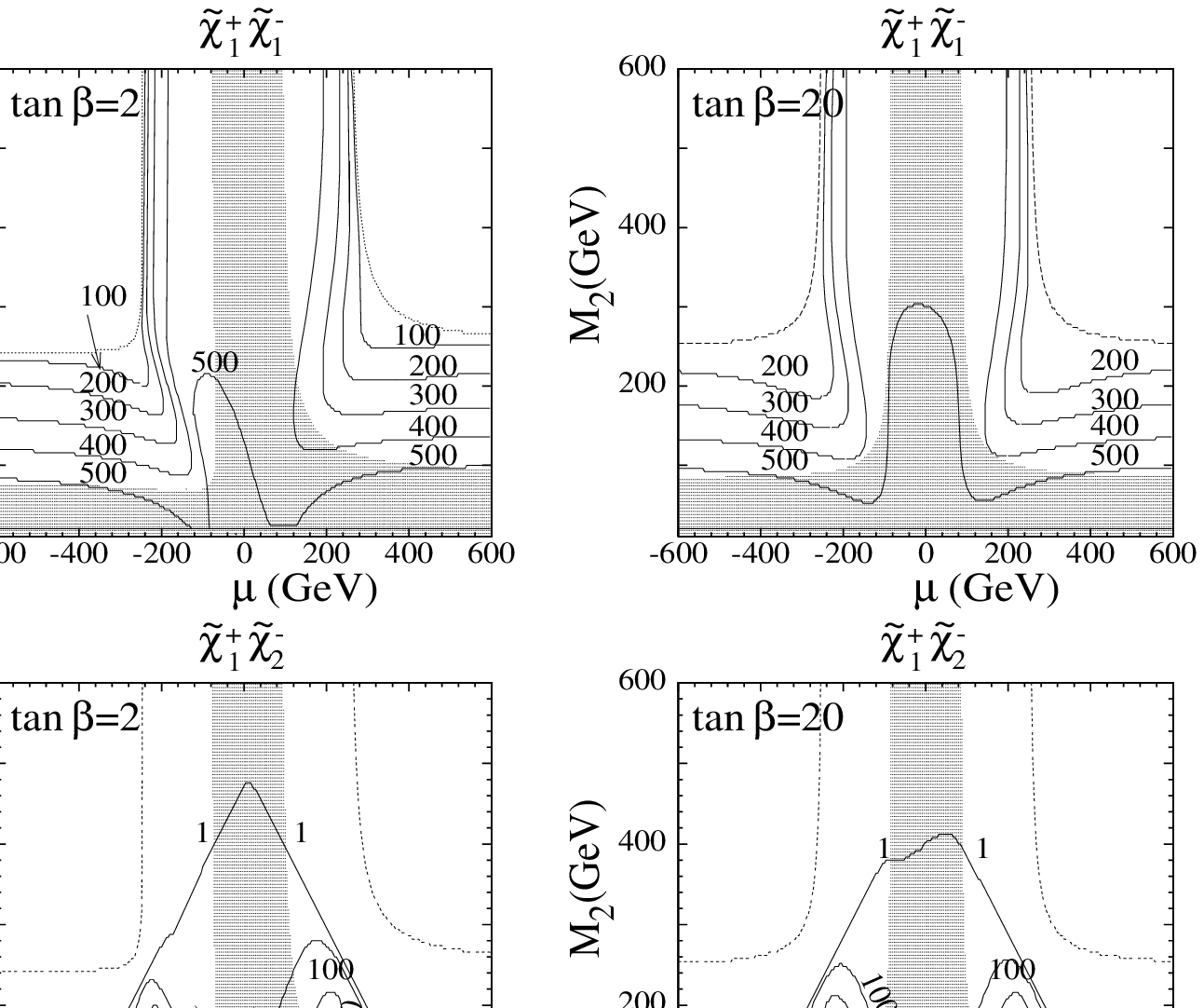}}
\end{center}
\end{figure}
\vspace{0.6in}
\noindent
{\bf Figure 5.}~{\it Contours of cross section (marked in
fb) for production at the {\em NLC} of a
pair of charginos for $\tb = 2$ and 20 and $M_{\widetilde{e_L}} =
M_{\widetilde{e_R}} = 150$ GeV.  }
\vspace{0.1in}

In Fig.~5, we present contours for the cross section (in fb) for the
production of $\Cp_1 \Cm_1$ and $\Cp_1 \Cm_2$ final
states in the ($M_2,\mu$) plane for two values of $\tb$, namely,
$\tb = 2$ for the plots on the left and $\tb = 20$ for the
plots on the right. The left selectron is assumed to have a mass of
150 GeV. It is clear from the figure that production of even a
single heavy chargino state leads to considerable reduction in the
cross section. Even so, the cross section is relatively large in
pockets close to the LEP-excluded region, where it is about 15--20\%
of the cross section for production of a pair of lighter charginos.
As most of the selected points (A)--(E) lie in or around these
pockets of large cross section, we can regard our results, obtained
by neglecting the production of a $\Cpm_1 \Cmp_2$ pair, to have
uncertainties of this order. However, as we shall see later, 
this excess cross section will be distributed over a wealth of
possible final states, so that the actual uncertainties in estimating many of 
these will be considerably reduced.

\begin{figure}[h]
\begin{center}
\vspace*{4.8in}
      \relax\noindent\hskip -5.0in\relax{\includegraphics{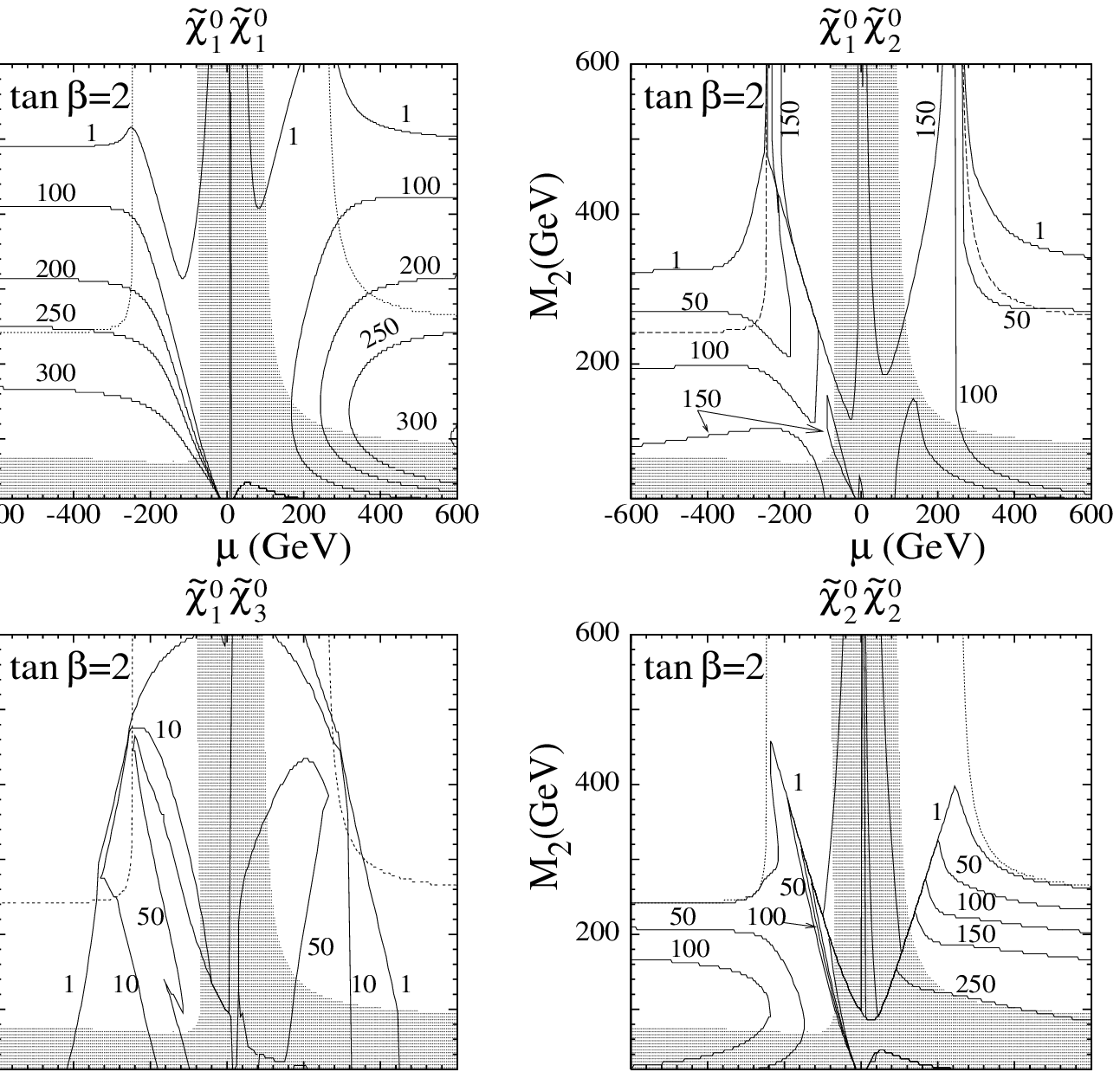}}
\end{center}
\end{figure}
\noindent
{\bf Figure 6.}~{\it Contours of cross section (marked in fb) for 
production at the {\em NLC} of a pair of 
neutralinos for $\tb = 2$ and 
$M_{\widetilde{e_L}} = M_{\widetilde{e_R}} = 150$ GeV.  }
\vspace{0.1in}

In Fig.~6, we present contours of cross section (in fb) for the
following pairs:  ($a$) $\N0_1 \N0_1$, ($b$) $\N0_1 \N0_2$, ($c$)
$\N0_1 \N0_3$, ($d$) $\N0_2 \N0_2$, all for $\tb = 2$. For this plot
the left and right selectrons are assumed to be degenerate, with the
mass being fixed at 150 GeV. Fig.~7 represents an identical plot
with $\tb = 20$. A glance at the cross sections for production of 
$\N0_3$ in association with an LSP, $\N0_1$, shows that the
production cross section is as severely suppressed (if not more),
compared to the lighter neutralino states as is the case of the
heavy chargino. Thus, we may feel justified in neglecting the
$\N0_3$ as well as the heavier $\N0_4$ and $\Cpm_2$ states. 
As before, this would lead to an error of 15--20\% at most,
which would again be spread out over a multitude of decay modes.

\begin{figure}[h]
\begin{center}
\vspace*{4.2in}
      \relax\noindent\hskip -5.0in\relax{\includegraphics{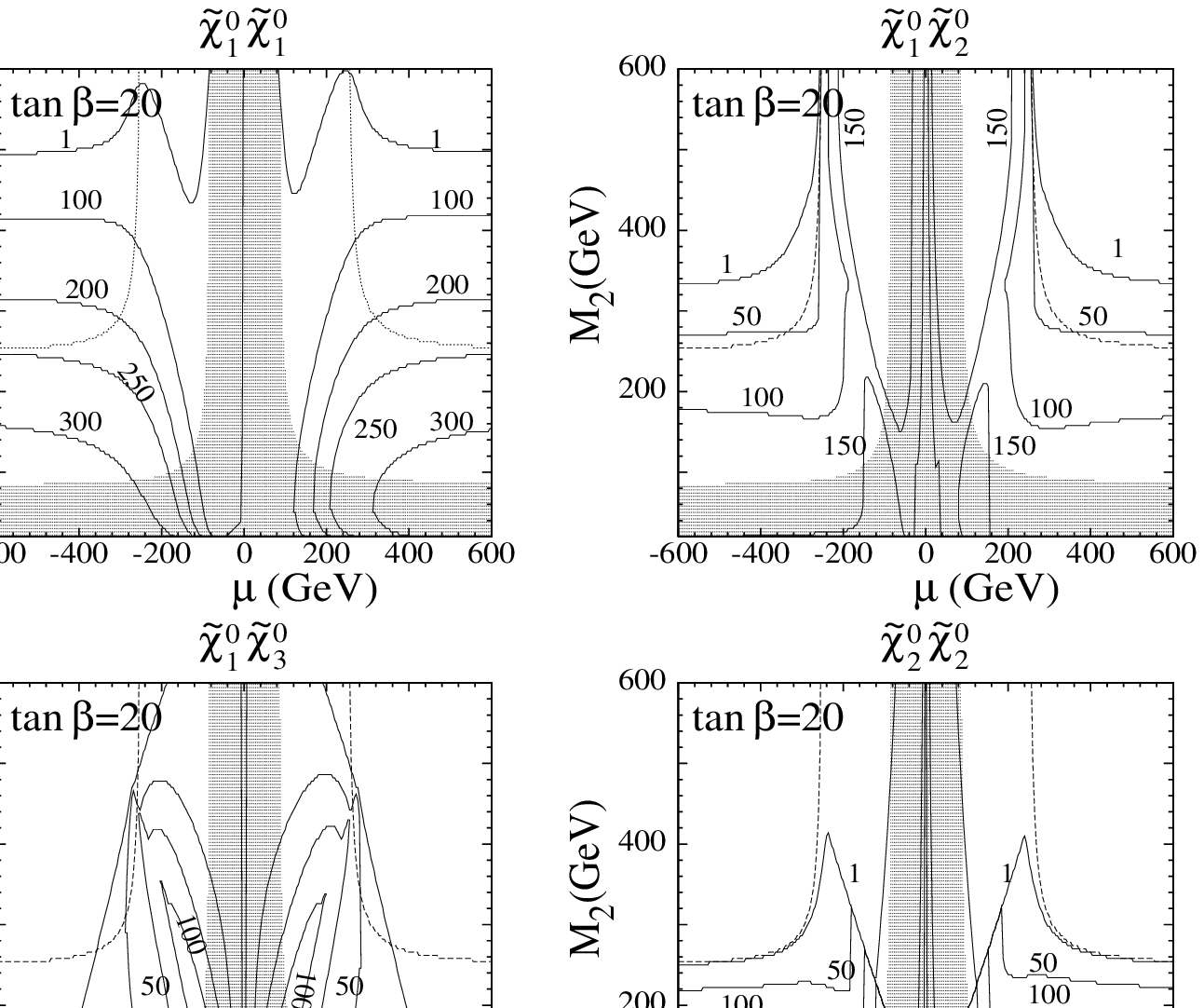}}
\end{center}
\end{figure}
\vspace*{0.55in}
\centerline{
{\bf Figure 7.}~{\it  Same as Fig.~6, but with $\tb = 20$.  } }
\vspace{0.1in}
 
Before concluding this section, some more general remarks are in
order. In the first place, the variation of the cross section with 
the mass of the
exchanged slepton as exhibited in Figs.~2 and 3 is
more-or-less straightforward to explain, given the composition of
the sparticles. The more complex variation of the cross section with
the parameters which make up the chargino and neutralino mass
matrices is shown in Figs.~5--7. These arise in a complicated way
through the diagonalization procedure and one cannot develop easy
physical arguments as to why the cross section is large or small.
The dark shading in Figs.~5--7 represents the region ruled out by
the chargino search at LEP-2 (running at 183 GeV) and corresponds to
the contour of (lighter) chargino mass $m_{\Cpm_1} < 91$ GeV. 
Searches for both \rp-conserving and \rp-violating 
supersymmetry at the four LEP
collaborations have essentially pushed the chargino mass to the 
kinematic limit\cite{LEP-Collaborations}, irrespective of the nature of the
\rp-violating operator. The broken line represents the
contours of chargino mass $m_{\Cpm_1} = 250$ GeV, which represents
the kinematic limit for chargino pair-production at the 500 GeV NLC.
Of course, the sample points (A)--(E) chosen for our analysis lie 
outside the LEP-2 ruled-out region.
 
For Figs.~5--7, we have chosen the selectron masses
to be $M_{\se_L} = M_{\se_R} = 150$ GeV. This is well above the
range of LEP-2, but it should be possible to pair-produce selectrons
of this mass at the 500 GeV NLC. Each of these selectrons can have
the following decays\cite{Kalinowski}: 
\bitem
\item to an electron and a neutralino; 
\item to a neutrino and a chargino (left selectron only); 
\item to a lepton and a neutrino (if there are $L$-violating operators of the
$LL \bar E$ form), or a pair of quarks (if there are $L$-violating
operators of the $LQ \bar D$ form). 
\eitem
Of these channels, in the
weak \rp-violation limit, we can neglect the third.  For a right
selectron, this means a unit branching ratio into electron and
neutralino. For a pair of left selectrons, we can have in the 
final states: 
$$\epem \N0\N0, \qquad 
   e^+ \nu \N0\Cm, \qquad 
   e^-\bar{\nu} \N0\Cp, \qquad 
  \nu \bar{\nu} \Cp\Cm, 
$$ 
where all possible neutralinos and charginos
allowed by kinematics will contribute.  The first channel usually
has the
largest branching ratio and will have essentially the same signals
as those which form the subject of this article, together with an
extra 
electron--positron pair with large energies and transverse momenta.
For the other signals, the chargino will decay through its gauge
couplings to $\Cpm_i \goes \N0_j f \bar f'$ where $f$ and $f'$ are
fermions differing by one unit in charge.  Thus we will have a pair
of neutralinos in the final state, accompanied by various
combinations of electrons, jets and neutrinos contributing to
missing energy. Another possibility which, in general, needs to be
considered\cite{Barger-Phillips-Keung}, 
is that of production of a pair of sneutrinos $ \snu_e $.
If the left selectron has a mass of 150 GeV, the mass of the
sneutrino $ \snu_e $ will be close to this, the relation between the
masses telling us that the sneutrino mass lies between 139--150 GeV,
depending on the value of $\tb$. These sneutrinos will certainly be
pair-produced in $\epem$ interactions at 500 GeV. The decay modes
will be analogous to those of the selectron, with similar final
states. A comprehensive discussion of all these signals is
desirable, but is beyond the scope of this work, which concentrates
on pair-production and decay of charginos and neutralinos. However,
we can guess that the signals and analysis will be very similar. 

For selectron masses above 250 GeV, it will not be possible to
produce selectron pairs at the 500 GeV NLC, and their effects will
appear solely in chargino and neutralino pair-production. The
sneutrino pair-production will be either suppressed or disallowed 
since the masses are close to or above the kinematic limit for somewhat
larger selectron masses. In this case, the only possibility for
detection of \rp~violation at the NLC will be through the signals
discussed in this article, though, it must be admitted that larger
selectron masses often lead to lower cross sections, as shown
in Figs.~2 and 3. However, as the following discussions will show,
a reduction in the cross sections by a factor as large as $2$ could still
yield substantial signals. 

\newpage 
\begin{center}
{\Large\bf 
               3. Chargino and Neutralino Decays
} \end{center}

In this article, we assume that the lightest neutralino is the LSP.
This is not forced upon us by cosmological arguments since we allow
the LSP to decay, but, as explained in the Introduction, 
it is a convenient hypothesis to make. 
The other assumption we make is that of all the \rp-violating
couplings, only one is dominant.  This is also an {\em ad hoc}
assumption, but it mimics the case of the SM Yukawa couplings, where
the top quark coupling is much larger than the others. In such a
scenario, it is possible to isolate signals arising from the LSP
decay, and this is what we analyze.

Once produced, the charginos and higher neutralinos will decay through 
various
channels into final states containing the LSP, which will then decay
into multi-fermion channels through \rp-violating couplings. Since
these are fairly well known\cite{Gaugino-production,Gaugino-decay}, 
we have not reproduced the formulae
for each of these decay modes but include a general discussion of all 
the channels.

\begin{figure}[h]
\begin{center}
\vspace*{4.2in}
      \relax\noindent\hskip -5.6in\relax{\includegraphics{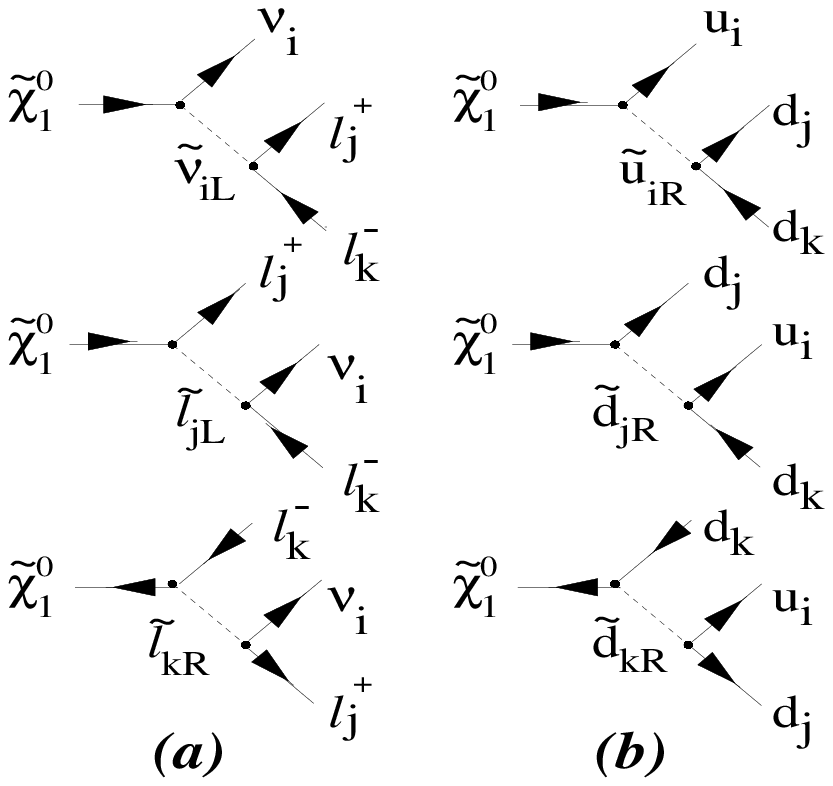}}
\end{center}
\end{figure}
\vspace{-1.4in}
\noindent
{\bf Figure 8.}~{\it Feynman diagrams contributing to the
decay of the {\em LSP} in the case of ($a$) $L$-violation with
$\l$ couplings, and ($b$) $B$-violation with $\lpp$ couplings.
The case of $\lp$ couplings can be obtained by replacing
$\nu_i \ell^+_j \ell^-_k \goes \nu_i d_j \bar d_k ~{\rm or}~
\ell^-_i u_j \bar d_k$ in ($a$).  }
\vspace{0.1in}

The most important decay of all these is, of course, the decay of
the LSP, which is the principal theme of this work.  This can occur
through the diagrams of Fig.~8($a$) and ($b$), which are mediated 
by sfermions. If the \rp-violating operator is of the
$LL \bar E$ type, the final states will have two charged leptons and
one neutrino, the flavours being determined by the single (dominant)
coupling involved. If the \rp-violating coupling is of the $LQ \bar
D$ type, the final states will have two quarks and one charged lepton, 
or two quarks and one neutrino. The relevant diagrams for the neutrino case
can be obtained from those in Fig.~8($a$) by replacing each charged 
lepton $\ell$ with a $d$-type quark. The charged lepton case can then be 
obtained from this by replacing the neutrino by a charged lepton and the 
$d$-type quark by a $u$-type quark. The LSP decay will result in 
either two hadronic jets accompanied by a charged lepton, or two
jets and missing momentum.  Finally, if the \rp-violating coupling
is of the $\bar U \bar D \bar D$ type, then the final state will generally 
consist of three hadronic jets. These modes are summarized
below, where $J$ denotes a hadronic jet.
$$
\N0_1
 \go \ell^+_1 \ell^-_2 + \mET~ ~~({\rm for}~~ LL\bar E), \qquad
\N0_1
 \go \ell^{\pm} JJ ~~{\rm or}~~ JJ + \mET~ ~~({\rm for}~~ LQ\bar D), \qquad 
\N0_1
 \go 3J ~~({\rm for}~~ \bar U \bar D \bar D),
$$

Formulae arising from the evaluation of these diagrams are presented
in Appendix A, where they agree with those already available in the
literature \cite{Baltz-Gondolo}. Each of these cases is discussed
separately in the following three sections, possible
signals and backgrounds being identified and estimated. Since the
LSP has only one decay mode, 
its branching ratio is always unity, and hence there is no
dependence on the actual magnitude of the \rp-violating coupling
involved. This is an important point, since it enables us to make
our analysis in the weak \rp-violation limit, and renders our
results robust in the sense that they are insensitive to the actual 
value of the \rp-violating
couplings involved. It is also worth drawing attention to the fact
that, since we regard the lightest neutralino as the LSP, there is
no possibility of sfermion {\rm resonances} in its decay.

When we come to the higher neutralinos and charginos, the situation 
becomes rather
complex, since all decays allowed by kinematic considerations can
occur. It follows that what may actually be expected to occur depends 
strongly on the
point in the MSSM parameter space. We must, therefore, consider, in
the general case, all possible decays of each chargino and/or neutralino. The 
decay of the heavier chargino to the lighter chargino occurs essentially
through the diagrams of Fig.~1($a$), where the $\epem$ pair can be
replaced by any pair of fermions.  The
decay of one neutralino to a different neutralino occurs through the
diagrams of Fig.~1($b$), where, once again, the $\epem$ pair is
suitably replaced. It is also possible for a chargino to decay into
a lighter neutralino or vice versa. Since all these decay channels have to
be considered, it is convenient to make a little summary of the 
possibilities, and this is presented below.
As before, we denote a hadronic jet by $J$.  It must be borne in
mind that the actual number of jets seen will not always be the same
as in the list below, since jets can be missed if they are soft or
if they merge into other jets. Similarly charged leptons will be
observed only if they pass all the acceptance cuts. 
$$
\begin{array}{lllll} 
\N0_2 & \go   & 
            \N0_1 \ell^+ \ell^-  & ~~{\rm or}~~\N0_1  + \mET        &
~~{\rm or}~~\N0_1 JJ              
\\
      & \hrar & 
             \Cpm_j \ell^\mp + \mET   & ~~{\rm or}~~\Cpm_j JJ ~~~(j = 1,2)    &    
\\ \\
\N0_3 & \go   & 
            \N0_i \ell^+ \ell^-  & ~~{\rm or}~~\N0_i  + \mET        & 
~~{\rm or}~~\N0_i JJ  ~~~(i = 1,2)
\\
      & \hrar & 
             \Cpm_j \ell^\mp + \mET   & ~~{\rm or}~~\Cpm_j JJ  ~~~(j = 1,2)   &
\\ \\
\N0_4 & \go   & 
            \N0_i \ell^+ \ell^-    & ~~{\rm or}~~\N0_i  + \mET          & 
~~{\rm or}~~\N0_i JJ  ~~~(i = 1,2,3)  
\nonumber \\
      & \hrar & 
             \Cpm_j \ell^+ + \mET      & ~~{\rm or}~~\Cpm_j JJ  ~~~(j = 1,2)    &
\\ \\
\Cpm_1 &\go   & 
            \N0_i \ell^\pm + \mET      & ~~{\rm or}~~\N0_i JJ  ~~~(i = 1,2,3,4) &     
\\ \\
\end{array}
$$
$$
\begin{array}{lllll} 
\Cpm_2 &\go   & 
            \Cpm_1 \ell^+ \ell^-    & ~~{\rm or}~~\Cpm_1  + \mET          &
~~{\rm or}~~\Cpm_1 JJ 
\\
       & \hrar &   
            \N0_i \ell^\pm + \mET    & ~~{\rm or}~~\N0_i JJ  ~~~(i = 1,2,3,4) & 
\\   
\end{array}
$$
This complex set of options is illustrated in Fig.~9, where a
typical mass spectrum is chosen and possible decays are
represented as transitions from a higher state to a lower. The left 
side
shows the most general case, when all possible cascade decays are
considered. It is straightforward to count that there are 15
possible decays when only the charginos and neutralinos are considered, 
and, if we
include the fact that either leptonic or hadronic final states can occur,
one has a total of 35 channels to consider. When this is combined
with the fact that a {\em pair} of charginos or neutralinos
 is produced in $\epem$
collisions, with all sorts of
combinations for the decays of each being possible, it is obvious
that a complete study is a gigantic task and well beyond the scope
of our present analysis. Fortunately, it proves possible to carry
out a meaningful analysis using only a subset of these production
and decay channels. The chief reason for this is that the heavier
neutralino and chargino states can be ignored in a first analysis,
as we have explained in the previous section. Moreover, 
it is well known that, for a
large part of the parameter space, the next-to-lightest neutralino
$\N0_2$ is nearly degenerate with the lighter chargino $\Cpm_1$, so
that its decay modes to the charginos (or vice versa) are
suppressed, if not forbidden. This certainly happens for the five
points where we have performed our analysis. 
A similar phenomenon occurs with the
heavier chargino and the heaviest neutralino. The corresponding 
decay modes have very little available phase space and 
may be ignored without much loss of generality.

\begin{figure}[h]
\begin{center}
\vspace*{4.2in}
      \relax\noindent\hskip -6.0in\relax{\includegraphics{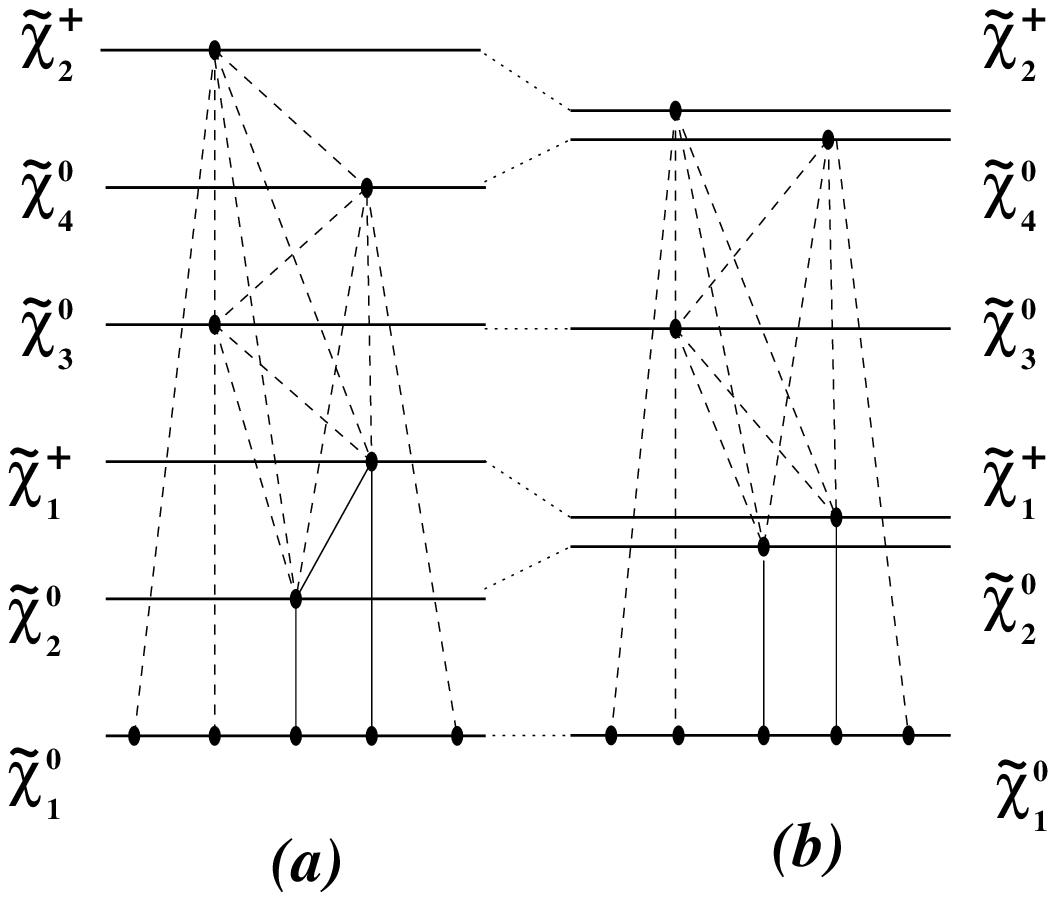}}
\end{center}
\end{figure}
\vspace*{-1.0in}
\noindent
{\bf Figure 9.}~{\it Schematic diagram showing all possible
cascade decays of higher charginos and neutralinos 
to the {\em LSP} in (a) the most
general case, and (b) a simplified case where two pairs of states 
are almost degenerate.  Broken lines indicate possible decays; solid
lines indicate those which have been considered in this article.}
\vspace{0.1in}

We now concentrate on the final states which could be of
interest at a 500 GeV $\epem$ collider. As we have explained in
detail in the previous section, these involve only the lowest three
states in Fig.~9, and correspond, on the left of Fig.~9 to the three
decays denoted by solid lines. If we take into account the
approximate degeneracy of the $\N0_2$ and the $\Cpm_1$, this is
reduced, on the right of Fig.~9, to just two decay modes. Of course,
these two lines really denote five decay modes, since $\N0_2 \goes
\N0_1 + (\ell^+\ell^- ~{\rm or}~ \nu\bar \nu ~{\rm or}~q\bar q$) and
$\Cp_1 \goes \N0_1 + (\ell^+\nu ~{\rm or}~~ q\bar q'$).  However,
this number is manageable, even when all the combinatorics are taken
into account, and we shall consider all these channels in the
subsequent analysis.

Our analysis is also limited to the case when the $t$-quark (either on shell
or off-shell) appear in the decays of the LSP, \ie the couplings 
$\lp_{i3k}$ and $\lpp_{3jk}$ do not form part of our analysis.
Our studies were done using a parton-level Monte Carlo event
generator, which does not yield high-quality results so far as
hadronic jets are concerned, but which does enable us to get a quick
and more-or-less correct picture of the signals that might be
visible at the NLC.


\vspace{0.3in}
\begin{center}
{\Large\bf 
    4. Signals from ${\bf L}{\bf L}\overline{\bf E}$ Operators 
} \end{center}

In this section, we concentrate on the signals that could be seen in
the case when the \rp~is violated by $LL\bar E$ couplings. As 
explained in the previous section, the LSP will decay into a pair of
charged leptons, with missing energy from a neutrino. Thus, if we
consider the basic process, \ie, a pair of LSP's being produced, then
the final state will have {\em four} charged leptons and missing
energy. The flavour of these leptons will be determined by the
coupling responsible and cannot be predicted \'a priori, unless the
coupling is known beforehand. If some of the heavier states 
are produced, then there will be additional leptons or jets
in the final state from cascade decays. We make a list of these
possibilities below and discuss each. 

For the production of a pair of LSP's, the final state will be
$ \N0_1 \N0_1 \go \ell^+_1 \ell^-_2 \ell^+_3 \ell^-_4 + \mET $, 
where, of the four charged leptons in the final state, two will be
of one flavour, and the other two will be of one (possibly different) 
flavour. Since there are two neutrinos contributing
to the missing energy, and each neutralino undergoes a three-body
decay, it will not be possible to reconstruct the mass of the LSP
from these final states. However, a final state with four hard
charged leptons and missing energy is sufficiently spectacular to
permit easy detection. There will be Standard Model backgrounds, of
course. The chief of these comes from $WWZ$ production ($W$-pair
production with a radiated $Z$), where all the gauge bosons decay
through leptonic channels. This background is unimportant at LEP-2,
which has insufficient energy to produce all the gauge bosons on-shell, 
but could, in principle, become significant at the 500 GeV NLC. 
It turns out, however, as we shall see, that this background 
is rather small when compared with the \rp-violating signals in the
multi-lepton channel. 

We now come to the issue of the production of higher neutralino and 
chargino states and their cascade decays to the LSP. 
We reiterate that this can lead to complicated signals, which 
are rendered more so by the fact that each higher chargino/neutralino
state can decay into a lower state and a gauge boson $W,Z$. The
latter can then decay
either leptonically or hadronically. For one neutralino decaying to
another, or the heavier chargino decaying into the lighter, we must
also distinguish between the cases of decay into a pair of charged
leptons and into a pair of neutrinos. Even at a single point in
parameter space, this makes for a somewhat messy analysis, 
although we limit the analysis to the low-lying states 
$\N0_1,\N0_2$ and $\Cpm_1$.

The signals which can be observed 
in the presence of $LL\bar E$ operators can be analyzed in the
following way\cite{Roszkowski}.  We consider the basic process
\be
\epem \goes \N0_1 \N0_1 \goes \ell^+_1 \ell^-_2 \ell^+_3 \ell^-_4 + \mET
~~(4\ell + \mET)
\ee
where two of the final state leptons are positively charged and two
are negatively charged. Assuming that the $\N0_2$ and the $\Cpm_1$
are nearly degenerate and thus their decay into one another may be
neglected (see above), 
the only cascade decays one needs to consider are those in
which a $\N0_2$ or a $\Cpm_1$ goes into a $\N0_1$ and a pair of
fermions. These processes then lead to the following signals.
$$
\begin{array}{lclcll}
\epem \goes   \N0_1 + \N0_2  
    & \go &   \N0_1 + \N0_1 \nu \bar \nu  
    & \goes &   4\ell + \mET
\\
    & \hrar & \N0_1 + \N0_1 \ell^+\ell^-  
    & \goes & 6\ell + \mET
\\
    & \hrar & \N0_1 + \N0_1 JJ            
    & \goes & 4\ell + \mET + 2~{\rm jets}
\\  \\
\epem \goes   \N0_2 + \N0_2 
    & \go   & \N0_1 \nu \bar \nu + \N0_1 \nu \bar \nu 
    & \goes & 4\ell + \mET
\\
    & \hrar & \N0_1 \ell^+\ell^- + \N0_1 \nu \bar \nu 
    & \goes & 6\ell + \mET
\\
    & \hrar & \N0_1 \ell^+\ell^- + \N0_1 \ell^+\ell^- 
    & \goes & 8\ell + \mET
\\
    & \hrar & \N0_1 \nu \bar \nu + \N0_1 JJ            
    & \goes & 4\ell + \mET + 2~{\rm jets}
\\
    & \hrar & \N0_1 \ell^+\ell^- + \N0_1 JJ 
    & \goes & 6\ell + \mET + 2~{\rm jets}
\\
    & \hrar & \N0_1 JJ + \N0_1 JJ 
    & \goes & 4\ell + \mET + 4~{\rm jets}
\\ \\
\epem \goes   \Cp_1 + \Cm_1 
    & \go   & \N0_1 \ell^+ \nu + \N0_1 \ell^- \bar \nu  
    & \goes & 6\ell + \mET 
\\
    & \hrar & \N0_1 \ell^+ \nu + \N0_1 JJ
    & \goes & 5\ell + \mET + 2~{\rm jets}
\\
    & \hrar & \N0_1 JJ + \N0_1 \ell^- \bar \nu  
    & \goes & 5\ell + \mET + 2~{\rm jets}
\\
    & \hrar & \N0_1 JJ + \N0_1 JJ
    & \goes & 4\ell + \mET + 4~{\rm jets}
\end{array} 
$$

This is not the end of the story, however, since every particle
produced in the final state is not always visible, given the
kinematic cuts and detector acceptances. We have chosen the following
criteria for observability of leptons/jets at a 500 GeV $\epem$ machine:
$$ 
p_T^\ell > 10~{\rm GeV} \ , \qquad |\eta_\ell| < 3 \ , \qquad 
p_T^J > 10~{\rm GeV} \ \qquad {\rm and} \qquad |\eta_J| < 3 \ . 
$$
Moreover, we assume that two partons with an angular separation of 
$ \delta R_{JJ} \equiv \sqrt{\delta \eta^2_{JJ} + \delta \phi^2_{JJ}} < 0.7 $
are merged into a single jet\footnote{We have checked that for our
parton-level analysis the Durham and Jade algorithms for jet merging
do not yield results substantially different from the cone algorithm 
used here.}.
Similarly, a lepton will be assumed to
be part of a jet if its angular separation $\delta R_{\ell J} < 0.4$.
These cuts represent educated guesses and may change a little when 
realistic simulations become available. However, we do not expect the 
qualitative features of our analysis to change too much even if a better 
set of kinematic cuts becomes available. 

Once all these criteria are applied, it is easy to see that signals
with large numbers of final state leptons and/or jets are likely to
get degraded to lower numbers of leptons and/or jets. Unobservable
leptons and/or jets would make (small) contributions to the missing
energy, augmenting the expectation from neutrinos produced in the
final state. Counting signals by the number of leptons, we 
classify the observable final states as having 0--8 leptons with or
without jets and always accompanied by substantial missing energy. To
get an idea of the strength of the signal available from the above
analysis, we show, in Table~3, the cross sections for these
signals for different pair-production modes (in the case of a
$\l$-coupling). We detail results for the five chosen points (A)--(E)
in the parameter space. For each point, there are 13
different configurations: these correspond to the cases \\
($a$)
$\N0_1 \N0_1$ pair production, where both LSP's decay to neutrino
plus dilepton (1 configuration), \\
($b$) $\N0_1 \N0_2$ pair production,
where $\N0_2$ decays to $\N0_1$ $+$  neutrinos, dilepton or
jets (3 configurations), \\
($c$) $\N0_2 \N0_2$ pair production, where
both $\N0_2$'s decay to $\N0_1$ $+$ neutrinos, dilepton or jets
(6 configurations), \\
($d$) $\Cp_1 \Cm_1$ pair production, where either
chargino can decay to $\N0_1$ plus leptons or jets (3
configurations). \\
After degradation, each of these makes a
contribution to the signal with $n$ leptons and missing energy (with
or without jets). These contributions are summed and displayed in Table 3. 
It is interesting that though the basic
process in this case starts with four leptons in the final state,
a large fraction of these gets degraded into states with three
(or even less) leptons. Fairly large cross sections may be obtained
for final states with 4 to 6 leptons in the final state. When
convoluted with a projected luminosity of about 10 fb$^{-1}$ at
the NLC, we see that these could yield some thousands of events
every year. Though signals with 7 or 8 leptons are quite rare,
given this large luminosity, we might expect to see a few of
these too.

The last column of Table~3 gives the Standard Model background for
each of these signals, where we have given the sum of
the principal backgrounds (see Appendix C). 
It is immediately obvious that final states with one or two leptons
have enormous backgrounds and cannot be used to look for 
\rp~violation. In any case, these arise, for the \rp-violating signal,
only as a result of degradation of signals with more leptons, and
are not of primary interest. 
On the other hand, once the signal has three or more leptons the
backgrounds are truly minuscule.  

One other question which can arise is what we expect from the
MSSM or other new physics\cite{Rizzo}
in case \rp~is conserved. In this case, instead of getting
$4\ell + \mET$ from the pair of LSP's, we will simply get missing 
energy. A rough estimate of the signals can be made by simply
mapping signals with $n$ leptons in Table~3 to signals with 
$n-4$ leptons. Keeping the backgrounds in mind, this means that
it should still be possible to see signals with three and four 
leptons, but generally at a lower level than what would be the case 
if \rp~is violated. The absence of signals with more than four
leptons would be a clear sign of 
\rp-conserving supersymmetry. Alternatively, one could say that
an unambiguous signal for \rp-violating supersymmetry would be 
large numbers of events with 4, 5 and 6 leptons. 

\newpage 

\footnotesize 
\vspace{-0.1in}
\begin{center}$$
\begin{array}{|c|l|cccc|c|c|}
\hline &
{\rm Signal}       
&  \N0_1 \N0_1  &  \N0_1 \N0_2  &  \N0_2 \N0_2  & \Cp_1 \Cm_1 &  
{\bf Total~Signal}~{\rm (fb)} & {\bf Background}~{\rm (fb)}    \\
\hline \hline {\bf A} 
& 1\ell + \mET &  ~~1.1~ & ~~0.4~ & ~~0.2~ & ~~1.5~   & ~~3.2~ & 8272.5
\\ \cline{2-8}
& 2\ell + \mET &  ~14.9~ & ~~5.2~ & ~~1.8~ & ~15.3~   & ~37.2~ & 2347.4 
\\ \cline{2-8}
& 3\ell + \mET &  ~91.7~ & ~25.3~ & ~~7.2~ & ~71.6~   & 195.8~ & ~~~1.5
\\ \cline{2-8}
& 4\ell + \mET &  212.8~ & ~49.6~ & ~13.6~ & 152.8~   & 428.8~ & ~~~0.4 
\\ \cline{2-8}
& 5\ell + \mET &  ~~0.0~ & ~37.8~ & ~19.3~ & 113.5~   & 170.6~ &  -
\\ \cline{2-8}
& 6\ell + \mET &  ~~0.0~ & ~39.6~ & ~21.6~ & ~26.9~   & ~88.0~ &  -
\\ \cline{2-8}
& 7\ell + \mET &  ~~0.0~ & ~~0.0~ & ~11.9~ & ~~0.0~   & ~11.9~ &  -
\\ \cline{2-8}
& 8\ell + \mET &  ~~0.0~ & ~~0.0~ & ~~8.0~ & ~~0.0~   & ~~8.0~ &  -
\\ \hline
\hline {\bf B} 
&1\ell + \mET  &  ~~0.2~ & ~~0.3~ & ~~0.3~ & ~~0.1~  &  ~~0.9~ & 8272.5
\\ \cline{2-8}
&2\ell + \mET  &  ~~4.5~ & ~~6.1~ & ~~5.6~ & ~~2.0~  &  ~18.1~ & 2347.4
\\ \cline{2-8}
&3\ell + \mET  &  ~34.7~ & ~38.0~ & ~28.7~ & ~17.0~  &  118.4~ & ~~~1.5
\\ \cline{2-8}
&4\ell + \mET  &  ~88.0~ & ~75.9~ & ~50.9~ & ~71.4~  &  286.2~ & ~~~0.4
\\ \cline{2-8}
&5\ell + \mET  &  ~~0.0~ & ~15.4~ & ~21.6~ & 148.5~  &  185.5~ &  -
\\ \cline{2-8}
&6\ell + \mET  &  ~~0.0~ & ~16.5~ & ~21.8~ & 124.2~  &  162.6~ &  -
\\ \cline{2-8}
&7\ell + \mET  &  ~~0.0~ & ~~0.0~ & ~~3.4~ & ~~0.0~  &  ~~3.4~ &  -
\\ \cline{2-8}
&8\ell +\mET   &  ~~0.0~ & ~~0.0~ & ~~2.3~ & ~~0.0~  &  ~~2.3~ &  -
\\ \hline 
\hline {\bf C} 
&1\ell + \mET   &~~0.3~   & ~0.2~  &~~0.6~  &~~0.5~   &~~1.6~ &  8272.5
\\ \cline{2-8}
&2\ell + \mET   &~~7.5~   & ~3.9~  &~~8.3~  &~~9.7~   &~29.4~ &  2347.4
\\ \cline{2-8}
&3\ell + \mET   &~56.3~   & 23.1~  &~39.1~  &~61.9~   &180.4~ &  ~~~1.5
\\ \cline{2-8}
&4\ell + \mET   &140.0~   & 44.9~  &~66.3~  &151.3~   &402.5~ &  ~~~0.4
\\ \cline{2-8}
&5\ell +\mET    &~~0.0~   & 11.6~  &~36.2~  &~93.6~   &141.4~ &   -
\\ \cline{2-8}
&6\ell + \mET   &~~0.0~   & 13.1~  &~37.6~  &~16.9~   &~67.6~ &   -
\\ \cline{2-8}
&7\ell +\mET    &~~0.0~   & ~0.0~  &~~7.3~  &~~0.0~   &~~7.3~ &   -
\\ \cline{2-8}
&8\ell +\mET    &~~0.0~   & ~0.0~  &~~5.3~  &~~0.0~   &~~5.3~ &   -
\\ \hline
\hline {\bf D} 
&1\ell + \mET  & ~~0.4  &   ~0.2 &   ~0.3 &  ~~0.2  &   ~~1.1  &  8272.5
\\ \cline{2-8}
&2\ell + \mET  & ~~7.8  &   ~3.8 &   ~5.0 &  ~~4.6  &   ~21.2  &  2347.4
\\ \cline{2-8}
&3\ell + \mET  & ~57.7  &   23.5 &   25.1 &  ~32.8  &   139.1  &  ~~~1.5
\\ \cline{2-8}
&4\ell + \mET  & 144.3  &   46.8 &   42.4 &  100.2  &   333.7  &  ~~~0.4
\\ \cline{2-8}
&5\ell + \mET  & ~~0.0  &   ~4.6 &   ~9.2 &  112.4  &   126.2  &   -
\\ \cline{2-8}
&6\ell + \mET  & ~~0.0  &   ~5.6 &   10.1 &  ~41.1  &   ~56.8  &   -
\\ \cline{2-8}
&7\ell + \mET  & ~~0.0  &   ~0.0 &   ~0.7 &  ~~0.0  &   ~~0.7  &   -
\\ \cline{2-8}
&8\ell + \mET  & ~~0.0  &   ~0.0 &   ~0.6 &  ~~0.0  &   ~~0.6  &   -
\\ \hline
\hline {\bf E}
&1\ell +\mET  &  ~~0.2~  &  ~0.1~ &  ~0.1~ &  ~~0.0~  &  ~~0.4~ & 8272.5
\\ \cline{2-8}
&2\ell +\mET  &  ~~5.1~  &  ~2.0~ &  ~1.4~ &  ~~1.7   &  ~10.2~ & 2347.4
\\ \cline{2-8}
&3\ell +\mET  &  ~45.9~  &  16.0~ &  11.6~ &  ~17.6~  &  ~91.1~ & ~~~1.5
\\ \cline{2-8}
&4\ell +\mET  &  136.7~  &  42.8~ &  28.5~ &  ~66.1~  &  274.1~ & ~~~0.4
\\ \cline{2-8}
&5\ell +\mET  &  ~~0.0~  &  ~3.1~ &  ~4.4~ &  ~77.4~  &  ~84.9~ &  -
\\ \cline{2-8}
&6\ell +\mET  &  ~~0.0~  &  ~4.8~ &  ~6.3~ &  ~27.9~  &  ~38.9~ &  -
\\ \cline{2-8}
&7\ell +\mET  &  ~~0.0~  &  ~0.0~ &  ~0.3~ &  ~~0.0~  &  ~~0.3~ &  -
\\ \cline{2-8}
&8\ell +\mET  &  ~~0.0~  &  ~0.0~ &  ~0.3~ &  ~~0.0~  &  ~~0.3~ &  -
\\ \hline 
\end{array} $$
\end{center}
\normalsize
{\bf Table 3.} {\it Showing the contribution (in fb) of
different (light) chargino and neutralino
 production modes to multi-lepton signals at
the {\em NLC} in the case of $\l$ couplings. The last column shows
the SM background. }
\vspace{0.1in}

In addition to this table, we present, in Fig.~10($a$), the
distribution in missing energy corresponding to the 
multi-lepton cases above for the point (A) in the parameter space. 
As expected, the missing energy grows progressively
softer as more and more leptons are seen: this is partly because
these correspond to states with less neutrinos and partly because
there are fewer undetected leptons which might have contributed to 
the missing energy. It is clear that a cut of 20 GeV on the 
minimum missing energy would leave the signal(s) practically
unaffected and this can be used for background reduction. 

Similarly, in Fig.~10($b$), we present histograms showing
the distribution in
the number of jets corresponding to each of the leptonic signals.
Again, quite naturally, there are fewer jets as the number of observed
leptons grows larger.
The point (A) in the parameter space is chosen for these plots.
Though our construction and treatment of jets is rather primitive,
we do not expect the qualitative features of this figure to change
when a more sophisticated analysis is performed. 

\begin{figure}[h]
\begin{center}
\vspace*{3.5in}
      \relax\noindent\hskip -6.0in\relax{\includegraphics{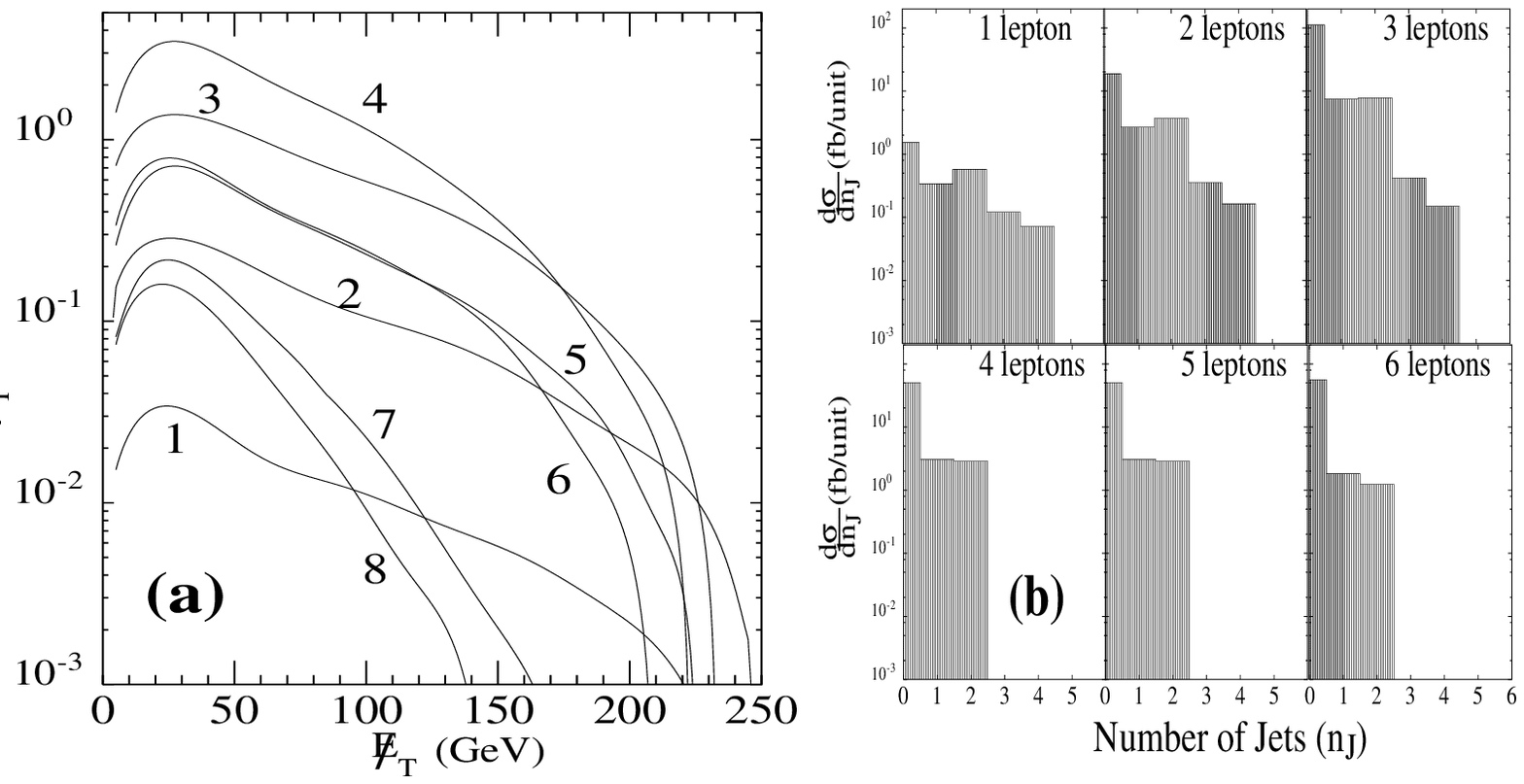}}
\end{center}
\end{figure}
\vspace{-0.5in}
\noindent
{\bf Figure 10.}~{\it Illustrating the distribution in ($a$) missing
transverse energy ($\mET$) and ($b$) the number of jets ($n_J$)
when the LSP decays through a $\l$ coupling. 
In ($a$) the
number of observable leptons is marked next to the relevant
curve.  }
\vspace{0.1in}

In Table 3, which is essentially illustrative, we have not considered
the particular $\l_{ijk}$ coupling, but presented a common value for
all flavour-combinations $ijk$. For better estimates, these must be 
appropriately convoluted with the respective detection efficiencies 
for $e,\mu$
and $\tau$ and also efficiency factors resulting from the different
isolation criteria (from hadronic jets) for a $\tau$ jet as compared to 
an electron or a muon track\cite{Ghosh-Raychaudhuri}. 
This part of the analysis has not been done since ours is
a preliminary study and we would like to obtain results of general
interest. Though it renders our analysis somewhat crude, we would
like to stress that the thrust of our work is in pinpointing the
crucial role of the heavier chargino and neutralino states in detection
of \rp-violating signals at the NLC, for which,we feel, this level of analysis 
is adequate. 

It is obvious that if \rp~is indeed violated through $LL\bar E$ operators
and the LSP (alone or in association with one of the 
higher states) is produced and decays at the NLC, then it should be
possible to see rather spectacular multi-lepton signals, even in one year of
running. This itself would be a major cause of excitement, but a
somewhat more tricky question would immediately arise: given that
some signals are seen, can we identify the specific coupling
responsible?  In pursuit of this answer, we note that the lepton content of
final states\footnote{
It is interesting that in the case of $LL\bar E$ couplings, the
Majorana nature of the neutralinos (both the LSP and the higher ones)
has no influence on the final states seen. This renders charge identification
irrelevant.} arising in LSP pair decay would be as follows:
$$
\begin{array}{clll} 
\l_{121}~:& 4e~,           & 2e + 2\mu~,    &  3e +\mu~;          \\
\l_{122}~:& 2e + 2\mu~,    & 4\mu~,         &  e + 3\mu~;        \\
\l_{123}~:& 2e + 2\tau~,   & 2\mu + 2\tau~, &  e + \mu + 2\tau~; \\
\l_{131}~:& 4e~,           & 2e + 2\tau~,   & 3e + \tau~;        \\
\l_{132}~:& 2e + 2\mu~,    & 2\mu + 2\tau~, &  e + 2\mu + \tau~; \\
\l_{133}~:& 2e + 2\tau~,   & 4\tau~,        &  e + 3\tau~;       \\
\l_{231}~:& 2e + 2\mu~,    & 2e + 2\tau~,   & 2e + \mu + \tau~;  \\
\l_{232}~:& 4\mu~,         & 2\mu + 2\tau~, & 2\mu + \tau~;      \\
\l_{233}~:& 2\mu + 2\tau~, & 4\tau~,        &   \mu + 3\tau~.
\end{array}
$$
Since all chargino and neutralino production modes are expected to end in 
a pair of
LSP's, which will then decay to the above flavour combinations, one
has to look at all multi-lepton states and identify the common
components. Thus, an exhaustive tabulation of the final states should
be able to tell us which flavour-combinations recur. From
this we can infer the coupling responsible, with at most a twofold
or threefold ambiguity. Though degradation of the number of leptons
will tend to obscure the issue, there should be enough 
four (or more)-lepton
final states produced directly through decay of the LSP to render
some sort of identification possible. We might then look to other
physics areas, perhaps precision measurement of lepton masses and
other similar parameters, to clinch the identification, and perhaps
extract the value of the coupling, which this analysis cannot give.


\vspace{0.3in}
\begin{center}
{\Large\bf 
    5.  Signals with ${\bf LQ}\overline{\bf D}$ Operators
} \end{center}

When we turn to the case of $LQ\bar D$ operators, the situation
becomes more complex. In the first place, as explained above, the LSP
has two decay modes, each with equal probability, but leading to
rather different signatures. The decay to a charged lepton and two
jets will lead, for $\N0_1 \N0_1$ production, to
a signal with an observable dilepton and (up to) four jets. The other
mode will lead to (up to) four jets with substantial missing energy and
momentum. One can also have a situation with one neutralino $\N0_1$ decaying
into either channel, which leads to a single charged lepton, missing
energy and (up to) four jets. This last signal will be enhanced by a
combinatoric factor of 2. Each signature will have large backgrounds.
However, in the first case, because of the Majorana nature of the
neutralinos, it may be possible to see {\em like-sign} dileptons (LSD's),
which have very little background from SM processes.

When cascade decays of the heavier chargino and neutralino states 
are included in our analysis, the situation becomes
even more complicated. If we merely consider their hadronic decays to the LSP, one
could see the above signals, but with an increased hadronic component. However,
the higher states can have leptonic decay modes too, and these will
interfere with or augment the leptonic signals, as the case may be. 
Accordingly, one has to analyze the different
signals with great care.

As in the previous section, we can analyze the signals in terms of
the various pair-production modes and the cascade decays. The basic
processes from pair-production of LSP's are
$$
\begin{array}{lclcccl}
\epem & \goes  &  \N0_1  + \N0_1  & \go   & \ell^\pm q \bar q' + \ell^\pm q \bar q'   
      & \go    &   2\ell + 4~{\rm jets}              \\
      &        &                  & \hrar &  \nu q \bar q  + \ell^\pm q \bar q' 
      & \go    &  \ell + \mET + 4~{\rm jets}         \\
      &        &                  & \hrar  &  \nu q \bar q + \nu q \bar q         
      & \go    &   \mET + 4~{\rm jets}                \\
\end{array}
$$

The first channel (which comprises a quarter of the signal) consists
of equal numbers of like-sign and unlike-sign dileptons. The signal
with unlike-sign dileptons will have large SM backgrounds especially
from $ZZ$, $t\bar t(g)$ and $\ell^+\ell^-Z$ channels (see Appendix C).
The signal with like-sign dileptons will, on the other hand, have 
practically no backgrounds and constitute a `smoking gun' signal for
the presence of Majorana neutralinos in the decay chain. It is well to
point out, however, that the only contributions to LSD signals will
come only when both leptons come from the decay of LSP's. Leptonic
decays of higher states to the LSP will never create LSD pairs,
though they may augment the LSD signal arising from the decay of
a $\N0_1 \N0_1$ pair. We shall discuss this issue in more detail
after enumerating the various decay modes and cross sections 
associated with the higher states. 

The second channel (which comprises half of the signal) consists of a
charged lepton, with missing energy and jets, where the lepton can be
of either sign. As shown in Appendix C, this will have huge SM backgrounds, 
with contributions from several channels. Isolation of \rp-violating signals 
from the background would require the careful application of
kinematic cuts. For example, the $t \bar t$ background can be simply
eliminated by requiring that no set of three jets should have an
invariant mass in the neighborhood of the known top quark mass. The
backgrounds with $W$'s and $Z$'s in the final state can be similarly handled 
if we require no pair of jets to reconstruct to the $W$
and $Z$ masses. However, it must be pointed out that with the 
usual uncertainties involved in jet reconstruction and the huge size
of the backgrounds, isolation of the signal may not be possible
over the {\it entire} parameter space. In fact, we may expect it to
be possible only when the signal has fairly large values, \eg,
at point (A) in the parameter space (see Table 4). 

The last channel, which requires one to trigger on 
missing energy, is not perhaps, the most rewarding mode to look for.
In the first place, it comprises just one-fourth of the signal.
Moreover, it may be mimicked by faulty reconstruction of jets, always
a possibility when fragmentation is taken into account. 
In our analysis, therefore, we have not taken this channel into account,
except in the case of higher state decays where it serves to augment 
the other multi-lepton signals.  

For the higher states, we restrict ourselves, once again, to the
$\N0_2$ and the $\Cpm_1$. Then, the processes of interest are
$$
\footnotesize
\begin{array}{lclclclll}  
\epem & \goes & \N0_1 + \N0_2 
& \go   & \N0_1 + \N0_1 \nu \bar \nu  & \go   & 2\ell + \mET + 4~{\rm jets} 
& ~~{\rm or}~~\ell + \mET + 4~{\rm jets}  
&                             \\
&     &                     
& \hrar & \N0_1 + \N0_1 \ell^+\ell^-  & \go   &  4\ell + 4~{\rm jets}
& ~~{\rm or}~~3\ell + \mET + 4~{\rm jets} 
&~~{\rm or}~2\ell + \mET + 4~{\rm jets}                              \\
&     &                     
& \hrar & \N0_1 + \N0_1 JJ            & \go   & 2\ell + 4~{\rm jets}
& ~~{\rm or}~~\ell + \mET + 4~{\rm jets}                                \\ 
\\
\epem & \goes & \N0_2 + \N0_2 
& \go   & \N0_1 \nu \bar \nu + \N0_1 \nu \bar \nu & \go & 
2\ell + \mET + 4~{\rm jets} & ~~{\rm or}~~ \ell + \mET + 4~{\rm jets} 
&       \\
&     &                     
& \hrar & \N0_1 \ell^+\ell^- + \N0_1 \nu \bar \nu & \go & 
4\ell + \mET + 4~{\rm jets} & ~~{\rm or}~~ 3\ell + \mET + 4~{\rm jets} 
&~~{\rm or}~~2\ell+\mET + 4~{\rm jets}     \\
&     &                     
& \hrar & \N0_1 \ell^+\ell^- + \N0_1 \ell^+\ell^- & \go & 
6\ell + 4~{\rm jets} & ~~{\rm or}~~ 5\ell + \mET + 4~{\rm jets} 
&~~{\rm or}~~4\ell+\mET + 4~{\rm jets}            \\
&     &                     
& \hrar & \N0_1 \nu \bar \nu + \N0_1 JJ  &  \go & 
2\ell + \mET + 6~{\rm jets} & ~~{\rm or}~~ \ell + \mET + 6~{\rm jets} 
&      \\
&     &                     
& \hrar & \N0_1 \ell^+\ell^- + \N0_1 JJ & \go  &
4\ell + 6~{\rm jets} & ~~{\rm or}~~ 3\ell + \mET + 6~{\rm jets} 
&~~{\rm or}~~2\ell+\mET + 6~{\rm jets}             \\
&     &                     
& \hrar & \N0_1 JJ + \N0_1 JJ & \go & 
2\ell + 8~{\rm jets} & ~~{\rm or}~~ \ell + \mET + 8~{\rm jets} 
&             \\
\\
\epem & \goes & \Cp_1 + \Cm_1 
& \go   & \N0_1 \ell^+ \nu + \N0_1 \ell^- \bar \nu  & \go & 
4\ell + \mET + 4~{\rm jets} & ~~{\rm or}~~ 3\ell + \mET + 4~{\rm jets}
&~~{\rm or}~~2\ell+\mET + 4~{\rm jets}      \\
&     &                     
& \hrar & \N0_1 \ell^+ \nu + \N0_1 JJ & \go &
3\ell + \mET + 6~{\rm jets} & ~~{\rm or}~~ 2\ell + \mET + 6~{\rm jets}
&~~{\rm or}~~\ell+\mET + 6~{\rm jets}      \\
&     &                     
& \hrar & \N0_1 JJ + \N0_1 \ell^- \bar \nu  & \go &
3\ell + \mET + 6~{\rm jets} & ~~{\rm or}~~ 2\ell + \mET + 6~{\rm jets} 
&~~{\rm or}~~\ell+\mET + 6~{\rm jets}     \\
&     &                     
& \hrar & \N0_1 JJ + \N0_1 JJ & \go & 2\ell + 8~{\rm jets}
& ~~{\rm or}~~ \ell + \mET + 8~{\rm jets} & .
\end{array} 
$$
\normalsize

As in the case of $LL\bar E$ couplings, the actual number of leptons
(and jets as well) 
will be degraded by kinematic cuts and detector acceptances.  In
Table 4 we set out the relative strengths of the various signals,
with different numbers of leptons accompanied (always) by jets and
(sometimes) by missing energy. One can again generate distributions
analogous to those of Fig.~10, but these have not been presented
here in the interests of brevity.

Just as we had 13 configurations in the case of $LL\bar E$ operators,
in the present case we can have 68 configurations. These correspond to \\
($a$)
$\N0_1 \N0_1$ pair production, with LSP's decaying to neutrino
or lepton plus jets (4 configurations), \\
($b$) $\N0_1 \N0_2$ pair production,
with the $\N0_2$ decaying to $\N0_1+$ neutrinos, dilepton or
jets (12 configurations), \\
($c$) $\N0_2 \N0_2$ pair production, with either
$\N0_2$ decaying to $\N0_1+$ neutrinos, dilepton or jets
(36 configurations), \\
($d$) $\Cp_1 \Cm_1$ pair production, with either
chargino decaying to $\N0_1+$ leptons or jets 
(16 configurations). \\
Each of these channels will make a contribution to the various 
multi-lepton signals. These are appropriately summed and displayed in Table 4. 

Let us take a closer look at these various signals.
The first channel, namely $\ell + \mET + {\rm jets}$, has
contributions from all the channels considered in Table 4 and this leads 
to a
large cross section, but it also has a  huge SM background 
and may not be the best option to look for $LQ\bar D$ couplings. 
The signal cross sections in Table 4 correspond typically to 
$2\sigma $ to $3\sigma$ fluctuations in the SM background. Mere observation 
of such an excess may not be enough to clinch the issue. However, it
may still be worth looking for this channel after applying cuts to
remove backgrounds where jets reconstruct to $W$ and $Z$ bosons, 
as discussed above. Unfortunately  these cuts are also 
likely to remove much of the contributions from higher chargino and 
neutralino states, which actually contribute the bulk of the signal
cross-section. It is not, therefore, clear that the mere observation
of an excess in the $\ell + \mET + {\rm jets}$ channel could be a 
signal for \rp-violation in the $LQ\bar D$ form. However, an excess in  
this channel could be used to supplement signals seen in other channels, 
which are discussed below.

The second channel, which contains a dilepton, is the best 
option for this kind of search. As before all chargino and neutralino
production channels will contribute to this signal and the total
cross section is reasonably high --- certainly greater than the
$4\sigma$ level for all the five points chosen for our analysis. 
This is the case irrespective of the charge of the leptons. 

\vspace*{-0.35in} 
\footnotesize
\begin{center}$$
\begin{array}{|c|l|cccc|c|c|}
\hline &
{\rm Signal}
&  \N0_1 \N0_1  &  \N0_1 \N0_2  &  \N0_2 \N0_2  & \Cp_1 \Cm_1
&  {\bf Total~Signal}~{\rm (fb)}  & {\bf Background}~{\rm (fb)} \\
\hline \hline {\bf A}
&1\ell + {\rm jets}  &   126.7  &  ~48.8  &  14.1  & 142.4  &  332.0 & 8262.5 \\ \cline{2-8}
&2\ell + {\rm jets}  &   ~27.9  &  ~47.2  &  21.6  & 128.4  &  225.1 & ~243.1 \\ \cline{2-8}
&3\ell + {\rm jets}  &   ~~0.0  &  ~28.0  &  21.8  & ~48.4  &  ~98.2 & ~~~1.5 \\ \cline{2-8}
&4\ell + {\rm jets}  &   ~~0.0  &  ~~6.0  &  14.2  & ~~6.5  &  ~26.7 & - \\ \cline{2-8}
&5\ell + {\rm jets}  &   ~~0.0  &  ~~0.0  &  ~6.3  & ~~0.0  &  ~~6.3 & - \\ \cline{2-8}
&6\ell + {\rm jets}  &   ~~0.0  &  ~~0.0  &  ~1.4  & ~~0.0  &  ~~1.4 & - \\ \cline{2-8}
\hline \hline {\bf B}
&1\ell + {\rm jets}  &   ~~8.1  &  ~26.3  &  30.8  & 185.7  &  250.9 & 8262.5 \\ \cline{2-8}
&2\ell + {\rm jets}  &   ~~0.1 &   ~15.3 &  ~22.7 &  106.1  &  144.2 & ~243.1 \\ \cline{2-8}
&3\ell + {\rm jets}  &   ~~0.0 &   ~~1.1 &  ~~5.0 &  ~~8.3  &  ~14.4 & ~~~1.5 \\ \cline{2-8}
&4\ell + {\rm jets}  &   ~~0.0 &   ~~0.0 &  ~~1.5 &  ~~0.2  &  ~~1.7 & - \\ \cline{2-8}
&5\ell + {\rm jets}  &   ~~0.0 &   ~~0.0 &  ~~0.1 &  ~~0.0  &  ~~0.1 & - \\ \cline{2-8}
&6\ell + {\rm jets}  &   ~~0.0 &   ~~0.0 &  ~~0.0 &  ~~0.0  &  ~~0.0 & - \\ \cline{2-8}
\hline \hline {\bf C}
&1\ell + {\rm jets}  &   ~14.7  &   18.2 &  ~46.0 &  124.1   &   203.0 & 8262.5 \\ \cline{2-8}
&2\ell + {\rm jets}  &   ~~0.3  &   13.6 &  ~44.2 &  ~26.8   &   ~84.9 & ~243.1 \\ \cline{2-8}
&3\ell + {\rm jets}  &   ~~0.0  &   ~1.2 &  ~11.7 &  ~~2.0   &   ~14.9 & ~~~1.5 \\ \cline{2-8}
&4\ell + {\rm jets}  &   ~~0.0  &   ~0.0 &  ~~4.5 &  ~~0.0   &   ~~4.5 & - \\ \cline{2-8}
&5\ell + {\rm jets}  &   ~~0.0  &   ~0.0 &  ~~0.4 &  ~~0.0   &   ~~0.4 & - \\ \cline{2-8}
&6\ell + {\rm jets}  &   ~~0.0  &   ~0.0 &  ~~0.0 &  ~~0.0   &   ~~0.0 & - \\ \cline{2-8}
\hline \hline {\bf D}
&1\ell + {\rm jets}  &  ~70.1  &   28.0  &  29.6 &  117.0  &  244.7 & 8262.5 \\ \cline{2-8}
&2\ell + {\rm jets}  &  ~~9.4  &   ~9.0  &  14.2 &  ~86.1  &  118.7 & ~243.1 \\ \cline{2-8}
&3\ell + {\rm jets}  &  ~~0.0  &   ~2.6  &  ~5.5 &  ~27.2  &  ~35.3 & ~~~1.5 \\ \cline{2-8}
&4\ell + {\rm jets}  &  ~~0.0  &   ~0.3  &  ~1.3 &  ~~3.1  &  ~~4.7 & - \\ \cline{2-8}
&5\ell + {\rm jets}  &  ~~0.0  &   ~0.0  &  ~0.2 &  ~~0.0  &  ~~0.2 & - \\ \cline{2-8}
&6\ell + {\rm jets}  &  ~~0.0  &   ~0.0  &  ~3.3 &  ~~0.0  &  ~~3.3 & - \\ \cline{2-8}
\hline \hline {\bf E}
&1\ell + {\rm jets}  &   ~66.2  &   23.4  &  17.0  &  70.2   &   176.8 & 8262.5 \\ \cline{2-8}
&2\ell + {\rm jets}  &   ~~9.8  &   ~7.6  &  ~8.0  &  43.6   &   ~69.0 & ~243.1 \\ \cline{2-8}
&3\ell + {\rm jets}  &   ~~0.0  &   ~2.2  &  ~2.6  &  26.8   &   ~31.6 & ~~~1.5 \\ \cline{2-8}
&4\ell + {\rm jets}  &   ~~0.0  &   ~0.3  &  ~0.8  &  12.5   &   ~13.6 & - \\ \cline{2-8}
&5\ell + {\rm jets}  &   ~~0.0  &   ~0.0  &  ~0.2  &  ~0.0   &   ~~0.2 & - \\ \cline{2-8}
&6\ell + {\rm jets}  &   ~~0.0  &   ~0.0  &  ~0.0  &  ~0.0   &   ~~0.0 & - \\
\hline
\end{array} $$
\end{center}
\vskip -5pt 
\normalsize
{\bf Table 4.}~{\it Showing the contribution (in fb) of
different (light) chargino and neutralino production modes to 
multi-lepton signals at
the {\em NLC} in the case of $\lp$ couplings. The last column shows
the SM background. }

We now come to the important issue of LSD signals. We have already
explained that one-eighth of the $\N0_1 \N0_1$ signal will consist of
LSD's: this means 35, 0.1, 0.4, 12, 12 LSD events at the points
(A)--(E) respectively, assuming an integrated luminosity of 10 fb$^{-1}$
per year. These numbers assume 100\% efficiency in charge identification,
whereas a figure of around 70-80\% would probably be more realistic.
The strong variation in these numbers over the parameter space
seems to indicate that one cannot confidently predict LSD signals
everywhere. However, the actual situation is not so bad, because
there will still be contributions from the cascade decays.

We can actually make a rough estimate of the contribution to LSD signals
from cascade decays using Table 4 and assuming that the decay widths of 
the $\N0_2$ and $\Cpm_1$ are dominated 
by the $Z$ and $W$-exchange channel respectively. This happens for a 
reasonably large part of the parameter space. In this case, we
estimate that the actual numbers of LSD events with 10 fb$^{-1}$ 
luminosity\cite{NLC-Report} 
will be around 180, 10, 75, 83 and 53 respectively for the
points (A)--(E). If we recall that there is practically no SM background
then, even with a charge detection efficiency of around 70\%, we
should see quite healthy numbers of like-sign dileptons in a year's running
of the NLC. 

Signals with three or more leptons will, in general, have small SM
backgrounds and should be seen without difficulty if the higher
charginos and neutralinos have a significant branching ratio to leptons. 
This is
certainly true of the points (A)--(E), but there
do exist points in the parameter space where the higher states decay
principally to the LSP accompanied by hadronic jets. In such cases,
the multi-lepton signals will be suppressed, but we should get a larger
{\it fraction} of LSD events. There is, thus, a trade-off, when we
look at all the signals. 

\begin{figure}[h]
\begin{center}
\vspace*{4.0in}
      \relax\noindent\hskip -6.0in\relax{\includegraphics{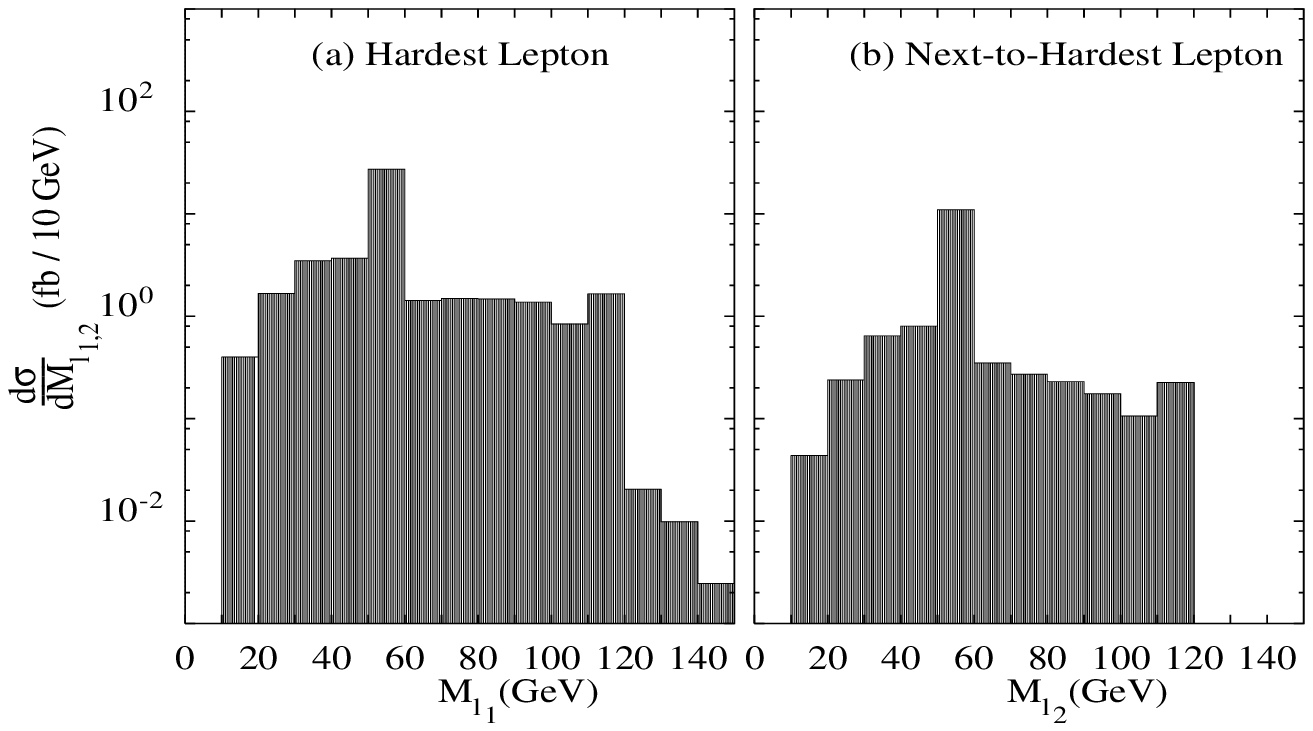}}
\end{center}
\end{figure}
\vspace{-1.5in}
\noindent
{\bf Figure 11.}~{\it Illustrating the distribution in invariant
mass reconstructed from ($a$) the hardest lepton and all jets in
the same hemisphere, and ($b$) the next-to-hardest lepton 
and all hadronic jets in its hemisphere, when all contributions
are summed over (signal only). The point {\rm (A)} in the parameter space 
is chosen.}
\vspace{0.1in}

To conclude this section, we make a few remarks on the possibility of
reconstructing the mass of the chargino and the neutralinos. 
A simple strategy is to focus on
the $n$-lepton-plus-jets signal, with or without missing energy. Assuming 
that the (highly relativistic) 
neutralinos are produced back-to-back in the laboratory frame, we
construct the invariant mass of the hardest lepton with
all jets in the same hemisphere, where the direction of the hardest 
jet is taken to be the hemisphere axis. The
invariant mass thus obtained should show a distinct peak at the
LSP mass. Cascade decays will, however, tend to broaden the distribution. 
In Fig.~11 we have shown
the invariant mass distribution from this construction using the
point (A) in the parameter space. This is done for ($a$) the hardest
lepton and ($b$) the next-to-hardest lepton (with all jets in its hemisphere). 
It is interesting to note 
that both the distributions show a classic resonance at the LSP mass 
(in the bin 50--60 GeV) as well as a secondary peak at the mass of the
next-to-lightest neutralino, or lightest chargino, both of which have
masses in the same bin 110--120 GeV. The secondary peak is easily 
identified as the bin beyond which the distribution exhibits a sharp fall.
If such a distribution is indeed seen above the background, we would be 
able to determine
simultaneously {\em three} sparticle masses, those of the $\N0_1, \N0_2, 
\Cpm_1$ respectively. Such information can then be used to determine 
two\footnote{To determine all three, we would also require to know the 
(small) mass-splitting between the $\N0_2$ and the $\Cpm_1$. This would
require a far better mass resolution than what we have assumed in this
work.} of the three 
free parameters $M_2$, $\mu$ and $\tb$. The measured cross-sections could
then be used to yield information on the remaining parameter as well as
the selectron masses. Much of this
would depend, however, on whether the distributions seen in Fig.~11
can be seen above the background. We do not expect any of the background
processes to show a strong resonant peak in this construction, but there
would be a large continuum. A detailed analysis, using a proper jet
reconstruction algorithm, needs to be performed, before this issue can
be properly resolved. We leave this issue to be settled in
a more focussed analysis in the future. 


\vspace{0.3in}
\begin{center}
{\Large\bf 
 6. Signals with 
    ${\bf \bar U}{\bf \bar D}{\bf \bar D}$ Operators
} \end{center}

The last case of \rp-violation is that of $\bar U \bar D \bar D$
operators with $\lpp$ couplings which arise in case baryon number is
violated. As discussed in section 3, in this case, the LSP will simply 
decay
into three hadronic jets. Thus the basic process with production of
$\N0_1 \N0_1$ will lead to a six-jet final state (with some jets lost
through merging). In itself, this is not impossible to study, but
there are two complications. The first is the usual case of higher
state production and decay, which can lead to leptons in the final
state, as well as (unwelcome) extra hadronic jets. The other is the
large SM background from $t\bar t$ production, which can lead to
six-jet final states if both the top quarks decay hadronically.
Other backgrounds, such as those from $WWZ$ production are not 
inconsiderable either.
Thus, kinematic strategies are of paramount importance in this case,
unlike the two previous cases, where a profusion of leptons in the
final state makes detection easy.

The basic process in this case is
\be
\epem \goes \N0_1 + \N0_1 
\goes u d d~(\bar u \bar d \bar d) + u d d~(\bar u \bar d \bar d)  
\goes 6~{\rm jets} 
\ee
where sets of three jets have invariant masses peaking at the
neutralino mass (assuming all the jets are seen) and the Majorana nature of
$\N0_1$ is taken into account. With the
introduction of cascades, we now have a large number of possibilities,
as in the previous two cases. These are listed below. 

$$
\begin{array}{lclclcl}
\epem & \goes & \N0_1 + \N0_2 
& \go   & \N0_1 + \N0_1 \nu \bar \nu & \go& 6~{\rm jets} + \mET \\
&&& \hrar & \N0_1 + \N0_1 \ell^+\ell^- & \go& 2\ell + 6~{\rm jets} \\
&&& \hrar & \N0_1 + \N0_1 JJ           & \go& 8~{\rm jets}
\end{array}
$$ $$ 
\begin{array}{lclclcl}
\epem & \goes & \N0_2 + \N0_2 
& \go   & \N0_1 \nu \bar \nu + \N0_1 \nu \bar \nu &\go& 6~{\rm jets} + \mET \\
&&& \hrar & \N0_1 \ell^+\ell^- + \N0_1 \nu \bar \nu &\go& 2\ell + 6~{\rm jets} + \mET \\
&&& \hrar & \N0_1 \ell^+\ell^- + \N0_1 \ell^+\ell^- &\go& 4\ell + 6~{\rm jets} \\
&&& \hrar & \N0_1 \nu \bar \nu + \N0_1 JJ           &\go& \mET + 8~{\rm jets}  \\
&&& \hrar & \N0_1 \ell^+\ell^- + \N0_1 JJ &\go& 2\ell + \mET + 6~{\rm jets}    \\
&&& \hrar & \N0_1 JJ + \N0_1 JJ           &\go& 10~{\rm jets}
\\ \\ 
\epem & \goes & \Cp_1 + \Cm_1 
& \go   & \N0_1 \ell^+ \nu + \N0_1 \ell^- \bar \nu  &\go& 2\ell + \mET + 6~{\rm jets} \\
&&& \hrar & \N0_1 \ell^+ \nu + \N0_1 JJ &\go& \ell + \mET + 6~{\rm jets} \\
&&& \hrar & \N0_1 JJ + \N0_1 \ell^- \bar \nu  &\go& \ell + \mET + 6~{\rm jets} \\
&&& \hrar & \N0_1 JJ + \N0_1 JJ &\go& 10~{\rm jets}
\end{array}
$$
Of course, final states with such large numbers of jets will
inevitably involve a great deal of jet merging, leading to states
with (much) lower jet multiplicity. A similar study involving
neutralino and chargino pair-production at LEP-1.5 has shown
\cite{Ghosh-Godbole-Raychaudhuri} that these lead dominantly to three- and four-jet final
states. Thus, the jet multiplicities shown above should not be taken
at all seriously. 

The final states arising from cascades could contain
one, two or even four leptons, as the case may be, and it might be
possible to trigger on these in a search. The states with one or two
leptons, as we have explained before, have
large SM backgrounds, though, as the table shows, it might not be
hopeless to look for an excess over the fluctuation in the background.
The three and four-lepton signals will be
easier to identify, but one must remember that they will come
accompanied by (up to) six jets. We, therefore, require these
leptons to satisfy isolation criteria to ensure that they are
not, indeed, part of the jets themselves. Following the earlier
sections, we present, in Table 5, a list of these multi-lepton
signals, always accompanied by multiple jets and (sometimes) by
missing energy.

In constructing this table, we consider a total of 17
configurations. These arise, like the previous ones from \\
($a$)
$\N0_1 \N0_1$ pair production, where each LSP decays to jets
(1 configuration), \\
($b$) $\N0_1 \N0_2$ pair production,
with $\N0_2$ decaying to $\N0_1~+$ neutrinos, dilepton or
jets (3 configurations), \\
($c$) $\N0_2 \N0_2$ pair production, with either $\N0_2$ decaying
 to $\N0_1~+$ neutrinos, dilepton or jets (9 configurations), \\
($d$) $\Cp_1 \Cm_1$ pair production, with either
chargino decaying to $\N0_1~+$ leptons or jets (4 configurations). \\
As before, all these are summed and displayed in Table~5. Each signal may or 
may not be accompanied by missing energy.

Of course, if the jet multiplicity is not taken into account,
then these signals will be rather similar to the (\rp-conserving)
MSSM backgrounds. This is because the leptons come entirely from
the decays of the chargino and higher neutralino states to the
LSP, irrespective of the decays of the LSP itself. Thus, in
order to isolate signals for $\lpp$ couplings, it is crucial to
study the multi-jet content of the final state.

\footnotesize
\begin{center}$$
\begin{array}{|c|l|cccc|c|c|}
\hline &
{\rm Signal}
&  \N0_1 \N0_1  &  \N0_1 \N0_2  &  \N0_2 \N0_2  & \Cp_1 \Cm_1 
&  {\bf Total~Signal}~{\rm (fb)} & {\bf Background}~{\rm (fb)} \\
\hline \hline {\bf A}
&1\ell + {\rm jets} &   ~~0.0  &  ~47.0  &  22.9 &  155.3  &  225.2 & 8262.5 \\ \cline{2-8}
&2\ell + {\rm jets} &   ~~0.0  &  ~38.1  &  26.2 &  ~30.7  &  ~95.0 & ~243.1 \\ \cline{2-8}
&3\ell + {\rm jets} &   ~~0.0  &  ~~0.0  &  12.5 &  ~~0.0  &  ~12.5 & ~~1.5\\ \cline{2-8}
&4\ell + {\rm jets} &   ~~0.0  &  ~~0.0  &  ~5.1 &  ~~0.0  &  ~~5.1 & - \\ \cline{2-8} \hline
\hline {\bf B}
&1\ell + {\rm jets} &   ~~0.0  &  ~21.3  & ~29.4 &  196.8  &  247.5 &  8262.5 \\ \cline{2-8}
&2\ell + {\rm jets} &   ~~0.0  &  ~12.3  & ~19.5 &  ~92.3  &  124.1 &  ~243.1 \\ \cline{2-8}
&3\ell + {\rm jets} &   ~~0.0  &  ~~0.0  & ~~3.2 &  ~~0.0  &  ~~3.2 &  ~~1.5 \\ \cline{2-8}
&4\ell + {\rm jets} &   ~~0.0  &  ~~0.0  & ~~0.9 &  ~~0.0  &  ~~0.9 & - \\ \cline{2-8} \hline
\hline {\bf C}
&1\ell + {\rm jets} &   ~~0.0  &   14.8  & ~44.3 &  115.0  &  174.1 &  8262.5 \\ \cline{2-8}
&2\ell + {\rm jets} &   ~~0.0  &   11.7  & ~39.9 &  ~16.4  &  ~68.0 &  ~243.1 \\ \cline{2-8}
&3\ell + {\rm jets} &   ~~0.0  &   ~0.0  & ~~7.7 &  ~~0.0  &  ~~7.7 &  ~~1.5 \\ \cline{2-8}
&4\ell + {\rm jets} &   ~~0.0  &   ~0.0  & ~~3.0 &  ~~0.0  &  ~~3.0 & - \\ \cline{2-8} \hline
\hline {\bf D}
&1\ell + {\rm jets}  &    ~~0.0 &  ~5.5   & 10.6  & 138.7  &  154.8 &  8262.5 \\ \cline{2-8}
&2\ell + {\rm jets}  &    ~~0.0 &  ~5.6   & 11.1  & ~45.0  &  ~61.7 &  ~243.1 \\ \cline{2-8}
&3\ell + {\rm jets}  &    ~~0.0 &  ~0.0   & ~0.8  & ~~0.0  &  ~~0.8 &  ~~1.5 \\ \cline{2-8}
&4\ell + {\rm jets}  &    ~~0.0 &  ~0.0   & ~0.4  & ~~0.0  &  ~~0.4 & - \\ \cline{2-8} \hline
\hline {\bf E}
&1\ell + {\rm jets} &   ~~0.0  &   ~3.5  &  ~4.8 &  ~90.2 &   ~98.4 &  8262.5 \\ \cline{2-8}
&2\ell + {\rm jets} &   ~~0.0  &   ~4.9  &  ~6.6 &  ~29.1 &   ~40.7 &  ~243.1 \\ \cline{2-8}
&3\ell + {\rm jets} &   ~~0.0  &   ~0.0  &  ~0.4 &  ~~0.0 &   ~~0.4 &  ~~1.5 \\ \cline{2-8}
&4\ell + {\rm jets} &   ~~0.0  &   ~0.0  &  ~0.3 &  ~~0.0 &   ~~0.3 & - \\ \hline
\end{array} $$
\end{center}
\normalsize
{\bf Table 5.}~{\it Showing the contribution (in fb) of
different (light) chargino and neutralino production modes to multi-lepton 
signals at the {\em NLC} in the case of $\lpp$ couplings.  }
\vskip 5pt

The question now arises: if we do indeed observe multi-jet signals,
with or without leptons and/or missing energy, what strategy should we 
adopt in order to identify whether these arise from
baryon-number-violating couplings, or from some other source?  It is
difficult to find a simple and all-encompassing answer. Perhaps the most 
straightforward
solution is to appeal to kinematics and carry out mass
reconstructions. Recognizing that the final state must always contain 
a pair of (generally) highly relativistic LSP's decaying into three jets 
apiece, we use the following strategy:
the hardest jet in the final state is identified and its invariant
mass is constructed with all the other jets lying in the same
hemisphere. Then, we look for the hardest jet in the opposite
hemisphere and again
construct its invariant mass with all other jets lying
in that hemisphere. We require these two invariant masses to lie
within 10 GeV of each other, noting that if they are equal, or nearly
equal, this is indicative of similar parentage. Once this criterion
is satisfied, we plot the two invariant mass distributions to see if they
show significant peaking. A similar strategy was followed by the
ALEPH Collaboration in their study of the (now-defunct) four-jet
anomaly \cite{ALEPH}. In an earlier study\cite{Ghosh-Godbole-Raychaudhuri} 
at LEP-1.5, it was shown
that a modest peak does indeed form. In the present
case, at the NLC, our results are illustrated in Fig.~12, for ($a$)
the hardest jet and ($b$) the hardest jet in the opposite hemisphere.
Once again, we have
chosen the point (A) in the parameter space. It is clear from the figure
that we have a clear resonance at the mass of the lightest
neutralino, with a sharp cutoff at the mass of the second neutralino
or (light chargino). However, this cutoff may not be clearly observable
because of the SM backgrounds, especially the one from $t \bar t$, which,
though it will peak around $m_t$, will have a non-vanishing tail in the
neighbourhood of 100--120 GeV. Backgrounds from $W$ and $Z$ bosons in the 
final state can be removed by requiring no two jets to have invariant
masses in the vicinity of $M_W$ and $M_Z$, though this cut has not actually been
applied in Fig.~12. Application of the cut would probably reduce the cross-section
somewhat, but make the resonance more sharp, since much of the smearing is due
to hadronic decays of the chargino and higher neutralino states, which take
place largely through $W$ and $Z$ exchanges. 

\begin{figure}[h]
\begin{center}
\vspace*{4.2in}
      \relax\noindent\hskip -6.0in\relax{\includegraphics{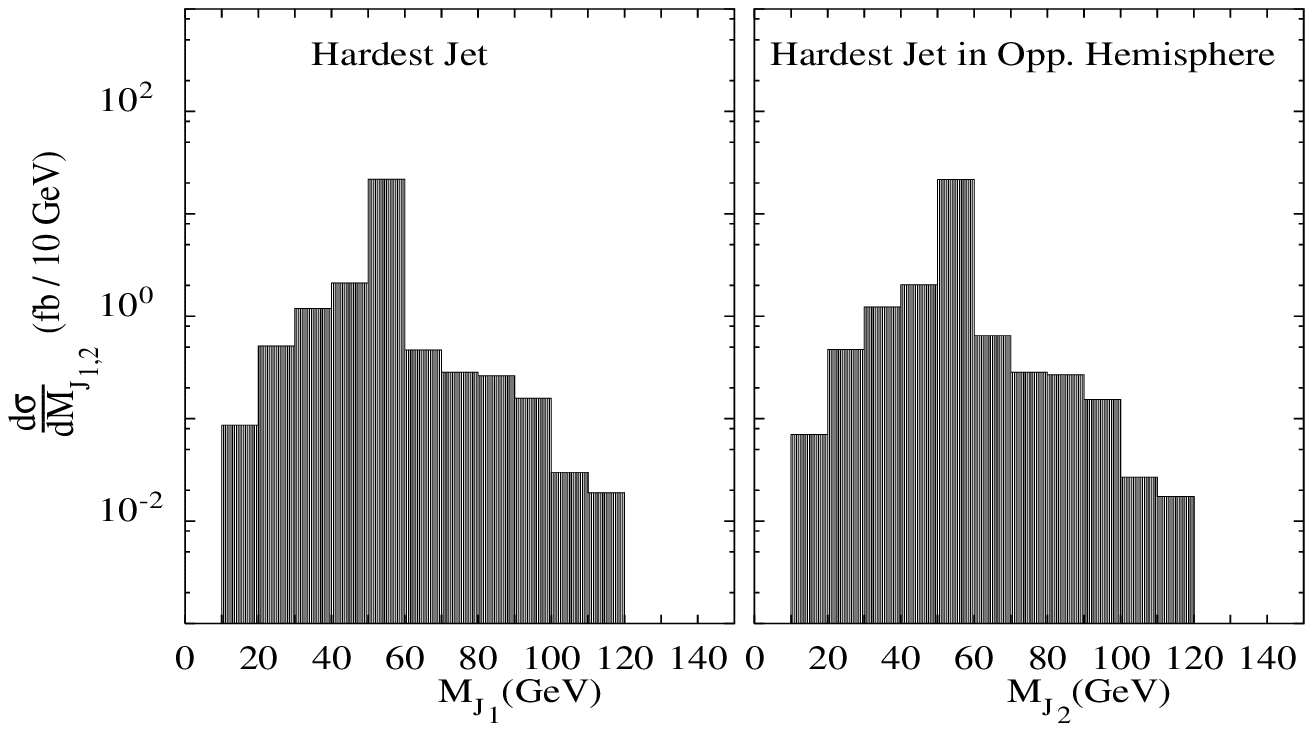}}
\end{center}
\end{figure}
\vspace{-1.4in}
\noindent
{\bf Figure 12.}~{\it Illustrating the distribution in invariant
mass reconstructed from ($a$) the hardest jet and all jets in
the same hemisphere, and ($b$) the hardest jet in the opposite hemisphere 
and all remaining hadronic jets, when all contributions
are summed over (signal only). The point (A) in the parameter space is chosen. }
\vspace{0.2in}

In case invariant mass distributions such as those shown in Fig.~12
are seen, it may be possible to determine the mass of the LSP, and
the (degenerate) $\N0_2$ and $\Cpm_1$. As explained in the previous
section, it might then be possible to determine some of the parameters 
of the MSSM; conversely, absence of the signals predicted here would
constrain the parameter space (assuming the presence of $\lpp$
couplings). 

Finally, we remark on the possibility of identifying the $\lpp$
coupling responsible for the signal seen. This must depend on
flavour-tagging, which is far more difficult for hadrons than for
leptons, and must be considered well-nigh impossible for the lighter
quark flavours. $b$ and $c$ quarks may be tagged with some reasonable
efficiencies, but, given the large number of jets and their large
merging probabilities, we think it highly unlikely that any
flavour-tagging worth the name can be achieved. It may be necessary
to look to other physics possibilities to determine the exact nature
of the coupling, assuming signals for a $\lpp$ coupling are seen.


\vspace{0.3in}
\begin{center}
{\Large\bf 
                      7. Summary and Conclusions
} \end{center}

It might be well to go over the main points of our analysis again 
before ending. We have discussed chargino and neutralino pair
production at a 500 GeV $\epem$ collider in the weak $R$-parity-violation
limit, where the production modes are solely through the 
$R$-parity-conserving couplings and the lightest supersymmetric particle (LSP)
decays purely through \rp-violating couplings. Our analysis shows that
it is sufficient to consider production of the two lightest neutralinos and 
of the lightest chargino. This simplifies matters considerably,
since the higher chargino and neutralino states can have very complicated 
cascade decays,
which would render the book-keeping almost an impossible task. We
have indicated how a non-observation of supersymmetry signals at
a 500 GeV machine would help us to make a very similar analysis at
a 1 TeV machine. 

We have then discussed the decays of the lightest neutralino (LSP)
through the three types of \rp-violating couplings assuming the top quark plays
no role in these decays. At the same time
we have considered decays of the (relevant) higher states to the
LSP, with accompanying fermions. These result in a wealth of final
state topologies, which we have classified by the number of hard 
isolated leptons. The observability of these leptons has been analyzed 
using a simple parton-level Monte Carlo event generator, which also
enables us to get approximate distributions in missing energy and the
jet-multiplicity in the final state. For $\l$-type couplings, the signal
is basically an excess of multi-lepton states, together with substantial
missing energy. For $\lp$-type couplings, the signal consists of leptons
and multiple jets in the final state, but we utilize a rather simple-minded 
algorithm to reconstruct the LSP and higher neutralino masses (within some 
resolution). This actually works pretty well and enables us to predict
that such signals, if seen, can be quite clearly identified as due to 
\rpv~supersymmetry. We also show that if charge identification of final-state 
leptons can be carried out with reasonable efficiency, then like-sign
dilepton (LSD) signals could be a very promising signal, insofar as they 
have practically no SM backgrounds. For $\lpp$ couplings, a similar mass 
reconstruction is
suggested: this shows a distinct resonance and could be again used to
indicate \rpv~supersymmetry. We have also discussed backgrounds and some 
simple strategies for their elimination, though a detailed simulation 
has not been attempted.

To conclude, then, 
we have made a detailed study of various possibilities for seeing the
\rp-violating version of the minimal supersymmetric Standard Model at
a 500 GeV linear $\epem$ collider. It is hardly necessary to point out that
one could make a very similar analysis at a muon collider of the same energy
and luminosity unless this is tuned to a Higgs resonance. Given the lack of any
concrete evidence of supersymmetry till date, and the fact that 
the $R$-parity-violating sector has very few theoretical constraints, 
the number of possible
scenarios for such signals is extremely large. A complete treatment
would be a massive task and probably not worth the effort at this early
stage, when a 500 GeV $\epem$ machine exists only on paper.
Nevertheless, we have isolated some of the important features of the
problem, and have identified some useful approximations that bring
a semblance of order to the multitude of cascade decays of
higher chargino and neutralino states, which can be produced with quite 
reasonable cross
sections at this energy. We hope that our work will encourage 
further studies of these extremely interesting signals.


\vspace{0.3in}
\begin{center}
{\large\bf   Acknowledgments.}
\end{center}

The authors are grateful to F.~Boudjema, J.~Kalinowski and B.~Mukhopadhyaya 
for discussions. DKG would like to thank the Centre for Theoretical Studies, 
Indian Institute of Science, Bangalore, where a large part of this work was 
done, for hospitality. RMG acknowledges partial financial support from the 
Department of Science and Technology (India), the National Science Foundation,
under NSF-grant-Int-9602567 and the Indo-French Collaboration no. 1701-1
{\it Collider Physics}. SR acknowledges the hospitality of the
Laboratoire de Physique Particles (LAPP), Annecy, where a part of this
work was done.

\newpage
\footnotesize

\def\pr#1,#2,#3 { {\em Phys.~Rev.}        ~{\bf #1},  #2 (19#3) }
\def\prd#1,#2,#3{ {\em Phys.~Rev.}        ~{\bf D#1}, #2 (19#3) }
\def\prl#1,#2,#3{ {\em Phys.~Rev.~Lett.}  ~{\bf #1},  #2 (19#3) }
\def\plb#1,#2,#3{ {\em Phys.~Lett.}       ~{\bf B#1}, #2 (19#3) }
\def\npb#1,#2,#3{ {\em Nucl.~Phys.}       ~{\bf B#1}, #2 (19#3) }
\def\prp#1,#2,#3{ {\em Phys.~Rept.}       ~{\bf #1},  #2 (19#3) }
\def\zpc#1,#2,#3{ {\em Z.~Phys.}          ~{\bf C#1}, #2 (19#3) }
\def\epj#1,#2,#3{ {\em Eur.~Phys.~J.}     ~{\bf C#1}, #2 (19#3) }
\def\mpl#1,#2,#3{ {\em Mod.~Phys.~Lett.}  ~{\bf A#1}, #2 (19#3) }
\def\ijmp#1,#2,#3{{\em Int.~J.~Mod.~Phys.}~{\bf A#1}, #2 (19#3) }
\def\ptp#1,#2,#3{ {\em Prog.~Theor.~Phys.}~{\bf #1},  #2 (19#3) }

\normalsize 

\newpage
\begin{center}
{\Large\bf   
                 Appendix A: Decay Widths of the LSP
} \end{center}

The decay of the LSP is mediated by \rp-violating couplings, which
arise from the superpotential of Eq.~(\ref{superpot}). In terms of
component boson and fermion fields, this leads to an interaction
Lagrangian for the lepton-number-violating part, which is
\bearr
{\cal L}_{int} = 
& - & \l_{ijk}
\left[ 
\sl^*_{Rk} \bar{\nu^c_i}  \frac{1-\gamma_5}{2} \ell_j
+ \sl^*_{Lj} \bar{\nu_i}    \frac{1+\gamma_5}{2} \ell_k
+ \snu^*_i   \bar{d_j}      \frac{1+\gamma_5}{2} \ell_k 
- (i \leftrightarrow j)
\right] 
\nonumber \\
& - & \lp_{ijk}
\left[ 
\sd^*_{Rk} \bar{\nu^c_i}  \frac{1-\gamma_5}{2} d_j
+ \sd^*_{Lj} \bar{\nu_i}    \frac{1+\gamma_5}{2} d_k
+ \snu^*_i   \bar{d_j}      \frac{1+\gamma_5}{2} d_k \right]
\nonumber \\ 
&   & \lp_{ijk}
\left[
  \sd^*_{Rk} \bar{\ell^c_i} \frac{1-\gamma_5}{2} u_j
+ \su^*_{Lj} \bar{\ell_i}   \frac{1+\gamma_5}{2} d_k
- \sl^*_{Li} \bar{u_j}      \frac{1+\gamma_5}{2} d_k 
\right] 
+ {\rm H.c.}
\eearr
Using the first line of this interaction Lagrangian, we can now calculate the 
decay
width of the neutralino in the case of a $\l_{ijk}$ coupling. We
present below complete formulae for the squared and spin-averaged
matrix element for the process $\N0_1(P) \go \nu_i(p_i) \ell^+_k(p_k)
\ell^-_j(p_j)$. The neutralino can also
decay into the $CPT$-conjugate of the right side because it is a
Majorana fermion. Thus we should include a factor of 2 when
calculating the decay width.)
\be
|{\cal M}(\N0_1 \go \nu_i \ell^+_k \ell^-_j)|^2 = 
\frac{12\pi\alpha\l^2_{ijk}}{\sin^2 \theta_W} 
\left[ T_{ii} + T_{jj} + T_{kk} + T_{ij} + T_{ik} + T_{jk} \right]
\ee
where the terms $T_{ii}$ etc. refer to the squares of the amplitudes
containing a sfermion of flavour $i$ in the propagator (see Fig.~8),
and $T_{ij}$ etc. are the interference terms. Similarly we denote the
masses of $\ell^\pm_i$ by $m_i$, the masses of $\sl_i$ by $\tm_i$ and
the mass of the neutralino by $M_1$. Then
\bearr
T_{ii} & = & \frac{p_j.p_k}{D_i^2} 
\left[ (G^2_{Li} + G^2_{Ri}) P.p_i 
\right]  \nonumber \\
T_{jj} & = & \frac{p_i.p_k}{D_j^2} 
\left[ 
(G^2_{Lj} + G^2_{Rj}) P.p_j  + 2 G_{Lj} G_{Rj} M_1 m_j
\right]  \nonumber \\
T_{kk} & = & \frac{p_i.p_j}{D_k^2} 
\left[ 
(G^2_{Lk} + G^2_{Rk}) P.p_k  + 2 G_{Lk} G_{Rk} M_1 m_k
\right]  \nonumber \\
T_{ij} & = &  \frac{-1}{D_i D_j}
\left[ 
G_{Li} G_{Lj} (-p_i.p_j P.p_k + p_i.p_k P.p_j + p_j.p_k P.p_i)
+ G_{Li} G_{Rj} M_1 m_j p_i.p_k 
\right]  \nonumber \\
T_{ik} & = & \frac{-\eta_1}{D_i D_k}
\left[ 
G_{Lj} G_{Lk} (p_i.p_j P.p_k - p_i.p_k P.p_j + p_j.p_k P.p_i)
+ G_{Li} G_{Lk} M_1 m_k p_i.p_j
\right]  \nonumber \\
T_{jk} & = &  \frac{-\eta_1}{D_j D_k}
\bigg[ 
G_{Lj} G_{Rk} (p_i.p_j P.p_k + p_i.p_k P.p_j - p_j.p_k P.p_i)
\nonumber \\
& & \hspace{1cm}
+ G_{Lj} G_{Rk} m_j m_k P.p_i
+ G_{Rj} G_{Lk} M_1 m_k p_i.p_j
+ G_{Rj} G_{Rk} M_1 m_j p_i.p_k
\bigg]  
\eearr
where $D_{i,j,k} = (s_{i,j,k} - \tm^2_{i,j,k})$, $s_{i,j,k} = (P -
p_{i,j,k})^2$, $\eta_1$ is the sign of the mass eigenvalue of the
lightest neutralino, and the couplings $G_{Li}, G_{Ri}$ etc. are
defined by the interaction of the neutralino with a lepton--slepton
pair:
\bearr
{\cal L}_{int} & = & 
- \frac{g}{2\sqrt{2}} \{ (1 + \eta_1) + i(1 - \eta_1) \}
\bar{\N0_i} 
\left\{ G_{Ln} \frac{1-\gamma_5}{2} 
      + G_{Rn} \frac{1+\gamma_5}{2} 
\right\} \ell_n \sl^*_n  + {\rm H.c.}
\eearr
The non-vanishing chiral couplings which enter into the above matrix element are
given by
\bearr
G_{Li} & = & \eta_1 ( N_{12} - N_{11} \tan \theta_W ),  
\nonumber \\
G_{Lj} & = & -\eta_1 ( N_{12} + N_{11} \tan \theta_W ),  
\nonumber \\
G_{Rj} & = & m_j N_{13} / (\sqrt{2} M_W \cos \beta), 
\nonumber \\
G_{Lk} & = & \eta_1 m_k N_{13} / (\sqrt{2} M_W \cos \beta), 
\nonumber \\
G_{Rk} & = & 2 N_{11} \tan \theta_W,
\eearr
where $N$ is the neutralino mixing matrix~(see Ref.~\cite{Rev_SUSY}).
In the above set of formulae, we note that we have not taken into
account the possibility of slepton resonances in the decay of the
neutralino. This is because the neutralino is considered to be the
LSP throughout this article\footnote{It is easy to translate the
above results to the case when there are slepton resonances. We
simply replace the propagators $(s_n - \tm^2_n)$ by the Breit-Wigner
forms $(s_n - \tm^2_n) + i \tm_n \Gamma_n$, where $\Gamma_n$ is the
total decay width of the $n$th slepton. This decay width is easily
calculated, since the slepton in question is now the LSP and will
undergo a two-body decay through the \rp-violating coupling
in question.}.
We have however, taken into account sign changes due to chiral rotations
of the neutralino fields necessitated by negative mass eigenvalues. 

It is trivial to translate the above set of results to the case of a
$\lp_{ijk}$ coupling, where the decay process is $\N0_1(P) \go
\nu_i(p_i) d_j(p_j) \bar{d}_k(p_k)$. We simply replace the
$\l$-coupling by the $\lp$-coupling and each lepton by the
corresponding $d$-type quark. The non-vanishing couplings of the 
neutralino to the
relevant quark--squark pairs are now given by
\bearr
G_{Li} & = & \eta_1 ( N_{12} - N_{11} \tan \theta_W ), 
\nonumber  \\
G_{Lj} & = & -\eta_1 ( N_{12} - \frac{1}{3} N_{11} \tan \theta_W ), 
\nonumber  \\
G_{Rj} & = & m_j N_{13} / (\sqrt{2} M_W \cos \beta), 
\nonumber \\
G_{Lk} & = & \eta_1 m_k N_{13} / (\sqrt{2} M_W \cos \beta), 
\nonumber \\
G_{Rk} & = & \frac{2}{3} N_{11} \tan \theta_W.
\eearr
In the other case for a $\lp_{ijk}$ coupling, the relevant process is
$\N0_1(P) \go \ell^-_i(p_i) u_j(p_j) \bar{d}_k(p_k)$. If we neglect
the charged lepton mass (which is a pretty good approximation for the
kind of neutralino masses discussed in this article), then one can
again use the above formulae with appropriate replacements, and
taking
\bearr
G_{Li} & = & -\eta_1 ( N_{12} + \frac{1}{3} N_{11} \tan \theta_W ),  
\nonumber \\
G_{Ri} & = & m_i N_{13}/ (\sqrt{2} M_W \cos \beta), 
\nonumber \\
G_{Lj} & = & \eta_1 ( N_{12} + \frac{1}{3} N_{11} \tan \theta_W ),  
\nonumber \\
G_{Rj} & = & m_j N_{14} / (\sqrt{2} M_W \sin \beta), 
\nonumber \\
G_{Lk} & = & \eta_1 m_k N_{13} / (\sqrt{2} M_W \cos \beta), 
\nonumber \\
G_{Rk} & = & \frac{2}{3} N_{11} \tan \theta_W.
\eearr
The real difference between the cases for $\l$ and $\lp$ couplings
arises when kinematics allows one to have a top quark in the final
state, in which case it is possible to have $m_j = m_t$, which makes
$G_{Rj}$ large. However, these have not been considered in this work.

We now pass on to the case of baryon-number-violating $\lpp$
couplings, where the interaction Lagrangian is given by
\bearr
{\cal L}_{int} =
& - & \lpp_{ijk}
\left[
\sd^*_{Rk} \bar{d_j}  \frac{1+\gamma_5}{2} u^c_i
+ \sd^*_{Rj} \bar{u_i} \frac{1-\gamma_5}{2} d^c_k
+ \su^*_{Ri}  \bar{d^c_j} \frac{1+\gamma_5}{2} d_k
\right]  + {\rm H.c.}
\eearr
This allows us to have the decay $\N0_1(P) \go u_i(p_i) d_j(p_j)
d_k(p_k)$ together with the $CPT$-conjugate of the right side. Using
the same kind of convention as in the case of
lepton-number-violating operators, the matrix element, squared and
summed over spins, has the form
\be
|{\cal M}(\N0_1 \go u_i d_j d_k|^2 = 
\frac{24\pi\alpha\lpp^2_{ijk}}{\sin^2 \theta_W} 
\left[ T_{ii} + T_{jj} + T_{kk} + T_{ij} + T_{ik} + T_{jk} \right]
\ee
where
\bearr
T_{ii} & = & \frac{p_j.p_k}{D_i^2}
\left[ (G^2_{Li} + G^2_{Ri}) P.p_i + 2 G_{Li} G_{Ri} M_1 m_i
\right]  \nonumber \\
T_{jj} & = & \frac{p_i.p_k}{D_j^2}
\left[ (G^2_{Lj} + G^2_{Rj}) P.p_j  + 2 G_{Lj} G_{Rj} M_1 m_j
\right]  \nonumber \\
T_{kk} & = & \frac{p_i.p_j}{D_k^2}
\left[ (G^2_{Lk} + G^2_{Rk}) P.p_k  + 2 G_{Lk} G_{Rk} M_1 m_k
\right]  \nonumber \\
T_{ij} & = &  \frac{-1}{D_i D_j}
 \bigg[
G_{Ri} G_{Rj} (-p_i.p_j P.p_k + p_i.p_k P.p_j + p_j.p_k P.p_i)
\nonumber \\
& & \hspace{1cm}
+ G_{Li} G_{Lj} m_i m_j P.p_k
+ G_{Li} G_{Rj} M_1 m_i p_j.p_k
+ G_{Ri} G_{Lj} M_1 m_j p_i.p_k
\bigg]
  \nonumber \\
T_{ik} & = & \frac{-\eta_1}{D_i D_k}
\bigg[ 
G_{Ri} G_{Rk} (p_i.p_j P.p_k - p_i.p_k P.p_j + p_j.p_k P.p_i)
\nonumber \\
& & \hspace{1cm}
+ G_{Li} G_{Lk} m_i m_k P.p_i
+ G_{Li} G_{Rk} M_1 m_i p_j.p_k
+ G_{Ri} G_{Lk} M_1 m_k p_i.p_j
\bigg]
  \nonumber \\
T_{jk} & = &  \frac{-\eta_1}{D_j D_k}
\bigg[ 
G_{Rj} G_{Rk} (p_i.p_j P.p_k + p_i.p_k P.p_j - p_j.p_k P.p_i)
\nonumber \\
& & \hspace{1cm}
+ G_{Lj} G_{Lk} m_j m_k P.p_i
+ G_{Lj} G_{Rk} M_1 m_k p_i.p_k
+ G_{Rj} G_{Lk} M_1 m_j p_i.p_j
\bigg]
\eearr
where we now take 
\bearr
G_{Li} & = & \eta_1 m_i N_{14}/ (\sqrt{2} M_W \sin \beta), 
\nonumber \\
G_{Ri} & = & -\frac{4}{3}\tan \theta_W N_{11}, 
\nonumber \\
G_{Lj} & = & \eta_1 m_j N_{13}/ (\sqrt{2} M_W \cos \beta), 
\nonumber \\
G_{Rj} & = & \frac{2}{3} \tan \theta_W N_{11}, 
\nonumber \\
G_{Lk} & = & \eta_1 m_k N_{13}/(\sqrt{2} M_W \cos \beta), 
\nonumber \\
G_{Rk} & = &\frac{2}{3}\tan \theta_W N_{11}.
\eearr

\newpage
\begin{center}
{\Large\bf
                 Appendix B: The 350 GeV Option
} \end{center}

It is, of course, feasible, and presumably planned, that a linear $\epem$
collider would first run at the $t \bar t$ production threshold of about 
350 GeV for some time, before it actually goes on to run at 500 GeV. The 
situation
regarding chargino and neutralino production is actually better in this case, 
as the dominant 
$s$-channel diagrams would be less suppressed, though of course, there
would be kinematic restrictions. The signals discussed in
this article can then be studied at this energy as well. However, the
SM backgrounds are also enhanced considerably. This is illustrated 
in Table 6, where we have presented 
the analogues of Tables 3--5, for the point (A) in the parameter space. 
A glance at the table will reveal that multi-lepton signals from $\l$
couplings will be trivially discernible over background, though one
and two-lepton signals will not. For the other two cases, the signals
are also quite promising. 

\footnotesize
\vspace{-0.1in}
\begin{center}$$
\begin{array}{|c|l|cccc|c|c|}
\hline &
{\rm Coupling}
&  \N0_1 \N0_1  &  \N0_1 \N0_2  &  \N0_2 \N0_2  & \Cp_1 \Cm_1 &
{\bf Total~Signal}~{\rm (fb)} & {\bf Background}~{\rm fb}  \\
\hline \l
& 1\ell + \mET &  ~~3.4~ &  ~~1.3~ & ~~0.3 &   ~~3.7   &   ~~~8.7 & 12929.4 \\ \cline{2-8}
& 2\ell + \mET &  ~36.1~ &  ~11.4~ & ~~3.0 &   ~34.9   &   ~~85.4 & ~3356.8 \\ \cline{2-8}
& 3\ell + \mET &  160.8~ &  ~40.9~ & ~~9.3 &   121.2   &   ~332.2 & ~~~~0.5  \\ \cline{2-8}
& 4\ell + \mET &  254.8~ &  ~64.1~ & ~16.5 &   189.1   &   ~524.5 & ~~~~0.1  \\ \cline{2-8}
& 5\ell + \mET &  ~~0.0~ &  ~51.9~ & ~21.0 &   115.3   &   ~188.2 & - \\ \cline{2-8}
& 6\ell + \mET &  ~~0.0~ &  ~39.4~ & ~19.4 &   ~23.7   &   ~~82.5 & - \\ \cline{2-8}
& 7\ell + \mET &  ~~0.0~ &  ~~0.0~ & ~11.0 &   ~~0.0   &   ~~11.0 & - \\ \cline{2-8}
& 8\ell + \mET &  ~~0.0~ &  ~~0.0~ & ~~5.3 &   ~~0.0   &   ~~~5.3 &- \\ \hline
\hline \lp
& 1\ell + {\rm jets} & 224.5 & ~58.7 & 12.3  &  165.6  &   ~461.1 & 12929.4 \\ \cline{2-8}
& 2\ell + {\rm jets} & ~76.5 & ~61.1 & 20.0  &  172.4  &   ~330.0 & ~~283.6 \\ \cline{2-8}
& 3\ell + {\rm jets} & ~~0.0 & ~46.6 & 26.6  &  ~70.6  &   ~143.8 & ~~~~0.5 \\ \cline{2-8}
& 4\ell + {\rm jets} & ~~0.0 & ~13.0 & 16.1  &  ~10.2  &   ~~39.3 & - \\ \cline{2-8}
& 5\ell + {\rm jets} & ~~0.0 & ~~0.0 & ~8.5  &  ~~0.0  &   ~~~8.5 & - \\ \cline{2-8}
& 6\ell + {\rm jets} & ~~0.0 & ~~0.0 & ~1.9  &  ~~0.0  &   ~~~1.9 & - \\ \hline
\hline \lpp
& 1\ell + {\rm jets} &  0.0 &  55.3  & 19.8  & 205.1   &  280.2  & 12929.4 \\ \cline{2-8}
& 2\ell + {\rm jets} &  0.0 &  61.4  & 27.5  &  43.5   &  132.4  & ~~283.6 \\ \cline{2-8}
& 3\ell + {\rm jets} &  0.0 &   0.0  & 14.6  &   0.0   &   14.6 & ~~~~0.5 \\ \cline{2-8}
& 4\ell + {\rm jets} &  0.0 &   0.0  &  7.5  &   0.0   &    7.5 & - \\ \hline 
\end{array} $$
\end{center}
\normalsize
{\bf Table 6.}~{\it Illustrating the contribution (in fb) of
different (light) chargino and neutralino  production modes to 
multi-lepton signals at
the {\em NLC} in the case of $\l,\lp$ and $\lpp$ couplings for a
linear collider running at $\sqrt{s} = 350$ GeV. The point {\em (A)}
in the parameter space has been chosen. }
\vspace{0.1in}

One interesting feature of a run at 350 GeV will be that the
background from $t \bar t$ production will be highly suppressed. As a
result, the invariant mass construction used in Fig.~12 would probably
be more certain to yield signals distinguishable above background.
Apart from this, there does not seem to be any major advantage, from
the point of view of studying \rp-violation, of a run at  $\sqrt{s} = 350$ GeV.

\newpage
\begin{center}
{\Large\bf
      Appendix C: Standard Model Backgrounds to Multilepton Signals 
} \end{center}

In order to obtain Standard Model backgrounds to the multi-lepton signals 
discussed in the text, we have considered several processes listed in Tables
7 and 8. These include $2 \go 2$ processes like $\epem \go W^+W^-$, 
$\epem \go ZZ$ and $\epem \go t \bar t$, and a host of $2 \go 3$ processes.
These were evaluated using the MadGraph package and the cross-sections
were calculated using a parton-level Monte Carlo generator. The cuts
mentioned in Section 4 were applied and the results were convoluted with the 
appropriate leptonic and hadronic branching ratios of the $W$ and $Z$ bosons. 

In Table 7, we present a list of the contributions to the multi-lepton plus
missing (transverse) energy states from various SM processes. For these,
we require a minimum of 20 GeV for the $\mET$ (see Section 4) and this 
essentially means 
that only processes which have a hard neutrino in the final state need be
considered. 
In the third row of this table, we have added contributions from
$\epem \goes t \bar t g$ to those from $\epem \goes t \bar t$, since
the final states will be very similar, except for jet multiplicity.
This increases the cross-section by about 10\%.

It is also important to point out that we have included only processes
which have
cross-sections greater than 1 fb (before convoluting with branching
ratios). For this reason, we have not considered processes like
$\epem \goes t \bar t Z$ and $\epem \goes ZZZ$, which could also,
in principle, yield multi-lepton final states.

\footnotesize
\vspace{-0.1in}
\begin{center}$$
\begin{array}{|l|l|l|l|l|}
\hline
{\rm Process}    
& \ell^\pm + \mET  & \ell^+\ell^- +\mET & 3\ell +\mET & 4\ell +\mET \\ \hline 
\hline
\epem \go W^+W^- 
& 2361.4~~(4250.5) & 560.7~(1009.3)     &~~~~-        &~~~~-\\ \hline
\epem \go ZZ     
&~~~~~~~~~-        & ~13.6~~~~~(23.1)   &~~~~-        &~~~~-\\ \hline
\epem \go t \bar t~(g) 
&~233.7~~~~~~(0.0) &~55.5~~~~~(0.0)     &~~~~-        &~~~~- \\ \hline
\epem \go W^+W^-Z 
&~~13.6~~~~~~~(4.5)&~~3.2~~~~~~(1.0)    & 1.5~(0.5)   & 0.4~(0.1) \\ \hline
\epem \go \epem Z
&~~~~~~~~~-        &~11.6~~~~(14.1)     &~~~~-        &~~~~- \\ \hline
\epem \go \ell^+\ell^-Z
&~~~~~~~~~-        &~~9.2~~~~~(15.2)    &~~~~-        &~~~~-\\ \hline
\epem \go e^+\nu_e W^- + e^-\bar \nu_e W^+ 
& 1564.0~~(1876.8) & 743.2~~~(891.8)    &~~~~-        &~~~~-\\ \hline
\epem \go \ell^+\nu_{\ell}W^- + \ell^- \bar \nu_{\ell}W^+ 
& 1940.0~~(2910.0) & 921.4~(1382.1)     &~~~~-        &~~~~-\\ \hline
\epem \go \nu_e \bar \nu_e Z 
&~~~~~~~~~-        &~~24.6~~~~(13.3)    &~~~~-        &~~~~-\\ \hline
\epem \go \nu_{\ell}\bar \nu_{\ell}Z 
&~~~~~~~~~-        &~~~4.4~~~~~(6.9)    &~~~~-        &~~~~-\\ \hline
\epem \go u \bar d W^- +\bar u d W^+ 
& 2159.8~~(3887.6) &~~~~~~~~~-          &~~~~-        &~~~~-\\ \hline
\hline
{\rm Total} 
& 8272.5~(12929.4) & 2347.4~(3356.8)    & 1.5~(0.5)   & 0.4~(0.1)\\
\hline
\end{array} $$
\end{center}
\normalsize
{\bf Table 7.}~{\it Illustrating the contribution (in fb) of
different SM processes to multi-lepton final states relevant for 
the {\em NLC}, running at 500~(350) GeV in the case of $\l$ couplings. 
In the 6th, 8th and 10th row, 
$\ell = \mu,\tau$. In the last row, the contribution is summed over the
first two generations. }
\vspace{0.1in}

It is clear that the only signal which can really contribute to
final states with $> 2$ leptons and $\mET$ is $\epem \go W^+W^-Z$, and this
has a rather small cross-section. This is the real reason why the 
signals from $\l$ couplings are so striking. 

In Table 8, we present a similar list of processes which lead to signals with
multi-leptons and jets. In this case, we do not put any cut on the 
missing energy, but we require the final state to have a minimum of
two jets. 

\footnotesize
\vspace{-0.1in}
\begin{center}$$
\begin{array}{|l|l|l|l|}
\hline
{\rm Process}    
& \ell^\pm + {\rm jets}  & \ell^+\ell^- + {\rm jets} & 3\ell + {\rm jets} \\ \hline 
\hline
\epem \go W^+W^- 
& 2361.4~~(4250.5)       &~~~~~~~~~-         &~~~~-        \\ \hline
\epem \go ZZ     
&~~~~~~~~~-              & ~47.6~~~~~(80.9)   &~~~~-        \\ \hline
\epem \go t \bar t~(g) 
&~233.7~~~~~~(0.0)       &~55.5~~~~~~~(0.0)     &~~~~-        \\ \hline
\epem \go W^+W^-Z 
&~~13.6~~~~~~~(4.5)      &~~4.0~~~~~~~(1.3)    & 1.5~(0.5)   \\ \hline
\epem \go \epem Z
&~~~~~~~~~-              &~82.0~~~(100.0)     &~~~~-        \\ \hline
\epem \go \ell^+\ell^-Z
&~~~~~~~~~-              &~32.2~~~~(53.4)     &~~~~-        \\ \hline
\epem \go e^+\nu_e W^- + e^-\bar \nu_e W^+ 
& 1564.0~~(1876.8)       &~~~~~~~~~-          &~~~~-        \\ \hline
\epem \go \ell^+\nu_{\ell}W^- + \ell^- \bar \nu_{\ell}W^+ 
& 1940.0~~(2910.0)       &~~~~~~~~~-          &~~~~-        \\ \hline
\epem \go u \bar d W^- +\bar u d W^+ 
& 2159.8~~(3887.6)       &~~~~~~~~~-          &~~~~-        \\ \hline
\epem \go q\bar q Z 
&~~~~~~~~~-                 &~21.8~~~~(48.0)     &~~~~-         \\\hline
\hline
{\rm Total} 
& 8262.5~(12929.4)       & 243.1~~(283.6)      & 1.5~(0.5)   \\ \hline
\end{array} $$
\end{center}
\normalsize
{\bf Table 8.}~{\it Illustrating the contribution (in fb) of
different SM processes to multi-lepton final states relevant for  
the {\em NLC}, running at 500~(350) GeV in the case of $\lp$ and $\lpp$ 
couplings. In the 6th and 8th row,
$\ell = \mu,\tau$, while in the 9th row, the contribution is summed over the
first two generations. In the last row, the flavour sum is carried over 
all quarks except the $t$-quark.  }
\vspace{0.1in}

As before, the only process which leads to $> 2$ leptons in the final
state is $\epem \go W^+W^-Z$, and this has a very small cross section. 
The single-lepton plus jets and dilepton plus jets final states, however,
which form the major signals for \rp-violation with $\lp$ and $\lpp$
couplings, have large backgrounds. Most of these come from final states
with $W$ bosons, and hence, the removal of states where jet-pairs
reconstruct to $W$-bosons could be a useful technique for background
reduction. 

Finally, we comment on effects of initial-state radiation (ISR) on the
processes under consideration. The signals for chargino and neutralino
production arise essentially from $s$-channel processes and, as such,
would be enhanced by ISR effects, even though these effects are rather
small. Many of the backgrounds, however, are essentially $t$-channel 
effects and these would be suppressed rather than enhanced by ISR. We thus 
feel that our analysis, which neglects ISR effects, is a conservative one.

\end{document}